\documentclass[11pt]{article}
\usepackage{amsmath,amssymb,mathrsfs,dsfont,eufrak}
\newcommand{\bea}{\begin{eqnarray}}
\newcommand{\eea}{\end{eqnarray}}

\newsavebox{\uuunit}
\sbox{\uuunit}
    {\setlength{\unitlength}{0.825em}
     \begin{picture}(0.6,0.7)
        \thinlines
        \put(0,0){\line(1,0){0.5}}
        \put(0.15,0){\line(0,1){0.7}}
        \put(0.35,0){\line(0,1){0.8}}
       \multiput(0.3,0.8)(-0.04,-0.02){10}{\rule{0.5pt}{0.5pt}}
     \end {picture}}

\numberwithin{equation}{section}
\begin{document}

\thispagestyle{empty}
{}

\begin{center}
{\bf\LARGE
On the sigma-model of deformed special geometry}

\vspace{10mm}

{\large
{{\bf Gabriel Lopes Cardoso$^+$ and } 
{\bf Alvaro V\'eliz-Osorio$^*$}}
\vspace{1cm}

{\it 
$^+$
Center for Mathematical Analysis, Geometry, and Dynamical Systems\\ [1mm] 
Departamento de Matem\'atica, 
Instituto Superior T\'ecnico
\\ [1mm]
1049-001 Lisboa, Portugal \\ [2mm] 
$^*$
Departamento de F\'isica, Instituto Superior T\'ecnico
\\ [1mm]
1049-001 Lisboa, Portugal \\ [2mm] 

}}

\end{center}
\vspace{8mm}

\begin{center}
{\bf ABSTRACT}\\
\end{center}

\noindent
We discuss the deformed sigma-model that arises when considering four-dimensional $N=2$ 
abelian vector multiplets in the presence of an arbitrary chiral background field. 
In addition, we allow for a class of deformations of special geometry by non-holomorphic terms.
We analyze the geometry of the sigma-model
in terms of intrinsic torsion classes.
We show 
that, generically, the deformed geometry is non-K\"ahler.  We illustrate our findings with an example.
We also express the deformed sigma-model in terms of the Hesse potential that underlies the real
formulation of special geometry.

\clearpage
\setcounter{page}{1}
\tableofcontents

%\clearpage

%%%%%%%%%%%%%%%%%%%%%%%%%%%%%%%%%%%%%%%%%%

\section{Introduction}

As is well known, $N=2$ supersymmetric abelian vector multiplets
in four dimensions contain complex scalar fields $X$ that give rise to a non-linear sigma-model whose target manifold
is a special-K\"ahler manifold.  The target-space geometry 
is called special geometry
\cite{Cecotti:1988qn,Strominger:1990pd}.  
In the context of global (or rigid) supersymmetry, special geometry is called affine special geometry, while
in local supersymmetry (i.e. when coupling to supergravity) special geometry is called projective special geometry.
In the holomorphic formulation of special
geometry, the target-space metric is encoded in a holomorphic function $F(X)$.
In the context of local supersymmetry,
$F(X)$ is required to be homogeneous of degree two, and the K\"ahler potential of the target-space
is expressed in terms of $F(X)$ as
\cite{deWit:1984pk}
\begin{equation}
  \label{eq:spec-K-pot}
  K(z,\bar z) = - \ln \left[
  \frac{\mathrm{i}\big(\bar X^I\, F_I - \bar F_{\bar I} \,
  X^I \big)}{\vert X^0\vert^2} 
        \right]\,.
\end{equation}
Here the index $I$ labels 
the vector multiplets (in supergravity it takes the values $I=0,1,\ldots,n$), and $F_I$ denotes
$F_I = \partial F(X) / \partial X^I$.  Due to the homogeneity property of $F(X)$, 
this K\"ahler potential depends only on the 
projective coordinates
$z^i=X^i/X^0$ (which are also called `special' holomorphic) and their complex conjugates, where $i=1,\ldots,n$.

Special geometry may also be formulated in terms of `special' real coordinates \cite{Freed:1997dp}.  In the context
of global supersymmetry,
the special-K\"ahler manifold
is a Hessian manifold.  The underlying Hesse potential is related by Legendre transformation to $F(X)$
\cite{Cortes:2001qd}. The Hesse potential also plays a fundamental role in the 
real formulation of projective special geometry 
\cite{Hitchin:2005uu,Alekseevsky:1999ts,Cortes:2001qd},
as emphasized recently in \cite{LopesCardoso:2006bg,Ferrara:2006at,Cardoso:2010gc,VandenBleeken:2011ib,Mohaupt:2011aa}.

In supersymmetric theories, parameters such as coupling constants can be viewed as background fields that are
set to constant values, so that supersymmetry is left intact. Background fields may also be used to generate
more general couplings of vector multiplets \cite{deWit:1996ix,deWit:1996ag}, for instance to the square of the Riemann tensor \cite{Bergshoeff:1980is}.  The coupling of background fields to vector multiplets
is restricted by supersymmetry, since background fields correspond to certain representations of supersymmetry.
In the following, we will consider the coupling of vector multiplets to so-called chiral background
fields, which we denote by $\hat{A}$.
The holomorphic function $F$ will then also depend on $\hat{A}$,
so that
now we have $F(X, \hat A)$. 
We will make use of the underlying superconformal approach to supergravity, in which the fields $X^I$ and $\hat A$ are subjected to local complex scale transformations. Rather than working with $X^I$ and $\hat A$, 
we will work with rescaled
fields $Y^I$ and $\Upsilon$ that are invariant under these local transformations.  
Accordingly, we will consider functions $F(Y, \Upsilon)$.  These then describe special geometry in the presence of
a chiral background field $\Upsilon$.

The abelian vector fields of the vector multiplets undergo electric/magnetic duality transformations
under which the electric field strengths and their magnetic duals are subjected to symplectic 
transformations.
The complex scalar fields $Y^I$ and the holomorphic derivatives $F_I = \partial F/\partial Y^I$ of the
function $F(Y, \Upsilon)$ are transformed accordingly, i.e. $(Y^I, F_I)$ is subjected to the same
symplectic transformation as the field strengths and their dual partners.  The background field $\Upsilon$ is inert
under these transformations \cite{deWit:1996ix}.

The presence of the chiral background field leads to a deformation of the non-linear sigma-model.  Namely, by setting
$\Upsilon$ to a constant value, $\Upsilon$ becomes a deformation parameter
 that affects the geometry of the target manifold. 
A concrete example is provided by taking the background field to describe the coupling of vector multiplets
to the square of the Riemann tensor, and 
by considering the associated sigma-model in a space-time that asymptotes to $AdS_2 \times S^2$,
with $\Upsilon$ set to the constant value $\Upsilon = - 64$.  This is the situation encountered in the context
of the quantum entropy function \cite{Sen:2008vm}, where the scalar fields asymptote to constant values, while
fluctuating in the interior of space-time \cite{Dabholkar:2010uh,Gupta:2012cy}.

It is known that gauge and gravitational coupling functions in $N=2$ theories receive non-holomorphic corrections
that are crucial to ensure that a given model has 
the expected
duality symmetries.
An early example thereof is provided by the computation of 
the moduli dependence of string loop corrections to gauge couplings in heterotic
string compactifications \cite{Dixon:1990pc}.  
These modifications, as well as considerations based on the duality invariance of black hole entropy formulae
\cite{LopesCardoso:1999ur,LopesCardoso:2004xf}, 
suggested that special geometry can be consistently 
modified
by a class of non-holomorphic deformations, whereby
the holomorphic function $F(Y, \Upsilon)$ is replaced by a non-holomorphic function
\cite{LopesCardoso:2006bg,Cardoso:2008fr,Cardoso:2010gc}
\begin{equation}
F(Y, \bar Y, \Upsilon, \bar \Upsilon) = F^{(0)}(Y) + 2 \mathrm{i} \, \Omega (Y, \bar Y, \Upsilon, \bar 
\Upsilon) \;,
\label{eq:F-0-Om}
\end{equation}
where $\Omega$ denotes a real (in general non-harmonic) function.  
The dependence on the chiral background is encoded in $\Omega$.  The previous formulation based on $F(Y, \Upsilon)$
is recovered by taking $\Omega$ to be harmonic. 

It was shown recently in \cite{Cardoso:2012nh} that the non-holomorphic deformations of special 
geometry described by \eqref{eq:F-0-Om} occur in a generic setting.  Namely, a theorem was presented stating
that a general point-particle Lagrangian ${\cal L}$ which depends on coordinates $\phi$ and velocities $\dot{\phi}$
can be reformulated in terms of complex coordinates $x = \tfrac12 ( \phi + \mathrm{i} \, \dot{\phi} )$ and a 
complex function $F(x, \bar x)$, such that the canonical coordinates 
$(\phi, \pi = \partial {\cal L}/\partial \dot \phi)$
coincide with (twice) the real part of $(x, F_x)$, where $F_x = \partial F(x, \bar x)/\partial x$.  The function
can be decomposed as in \eqref{eq:F-0-Om}, namely
$F(x, \bar x) = F^{(0)} (x) + 2 \mathrm{i} \Omega(x, \bar x)$, where $\Omega$ is real.  This reformulation
exhibits features that are analogous to those of special geometry.

Based on these observations, we are led to 
study the modified sigma-model that arises when replacing 
$F(Y, \Upsilon)$ by the non-holomorphic
function \eqref{eq:F-0-Om}.  Although it is not known whether there exists
a supersymmetric effective action based on 
\eqref{eq:F-0-Om}, the resulting sigma-model represents an extension of the usual sigma-model based on $F^{(0)}(Y)$
that has symplectic covariance built into it.
The latter is a key feature of $N=2$ systems \cite{deWit:1984pk,deWit:1996ix}.

As stated below \eqref{eq:F-0-Om}, the dependence on the chiral background is encoded in $\Omega$.
In the following, whenever we turn on the chiral background so that
$\Omega \neq 0$, we will refer to the resulting sigma-model as deformed sigma-model.
We introduce the 
deformed sigma-model in section \ref{sec:defsm}. It is coupled to supergravity, and 
given in \eqref{lag-sig-def} in terms of the fields $Y^I$ and $\Upsilon$.
As already mentioned, we set $\Upsilon$ to a constant value, thereby treating it as a deformation parameter.
We may consider the following limiting cases.  When taking $\Omega$ to be harmonic, we obtain a deformed sigma-model
that is encoded in $F(Y, \Upsilon)$, as mentioned above.  We will refer to this situation as the Wilsonian
case.  The presence of non-holomorphic terms in $\Omega$ implies a departure from the Wilsonian case.
We may also consider decoupling supergravity, and we will refer to this case as the rigid case.

We analyse the geometry of the deformed sigma-model in terms of intrinsic torsion classes following 
\cite{Chiossi:2002tw}.
To do so, we first express the
deformed sigma-model in terms of projective coordinates.
In the presence of a chiral background, the set of projective coordinates is given by
$z^i = Y^i/Y^0 \,,\, \Psi = \Upsilon/ (Y^0)^w$, where $w$ denotes the weight of the chiral background field
$\hat A$ under scale transformations.   The deformed sigma-model may also depend on the phase of $Y^0$,
which we denote by $R = \bar Y^0/Y^0$. 
When $\Omega =0$,
it is well known that 
the sigma-model target-space metric does not depend on $R$, i.e. the target-space metric has an isometry
associated with this angular variable.  When $\Omega \neq 0$, however, we will show that this is 
not any longer the case and that the target-space metric is non-K\"ahler, in general.
When written in projective coordinates, we find that the deformed sigma-model is given by \eqref{eq:lag-nonhol-final}.
It takes a rather simple form.  The first term is the one that is present in the rigid limit when decoupling
supergravity.  Its form represents a generalization of the one that enters in the 
extrinsic construction of special-K\"ahler manifolds given in \cite{Alekseevsky:1999ts}. The second term
arises when coupling the deformed sigma-model to supergravity.  We discuss the rigid limit in subsection 
\ref{sec:rigid-lim}.

When treating $\Upsilon$ as a deformation parameter by setting it to a constant value, 
the chiral background field $\Psi$ ceases to be an independent field.  It becomes expressed in terms of
$z^i, \bar z^i$ and $R$, and this in turn affects the geometry of the deformed target-space.
This effect is already present at the Wilsonian level, as we discuss in subsection \ref{sec:wils}.
In section \ref{sec:int-tors} we 
analyze the geometry
of the deformed target-space in terms of intrinsic torsion classes. 
We restrict to 
a hypersurface $R = {\rm constant}$ and take the number of vector multiplets to be specified by $n=3$, in order
be able to use the analysis given in \cite{Chiossi:2002tw}.  We work to first order in $\Omega$ or, equivalently,
to first order in the deformation parameter $\Upsilon$.  We express the intrinsic torsion in terms of 
target-space metric
coefficients that are of first-order in $\Upsilon$.  We find that the torsion class ${\cal W}_1$ vanishes,
whereas ${\cal W}_2$ is in general non-vanishing. For a manifold to admit an integrable complex structure, both
${\cal W}_1$ and ${\cal W}_2$ have to vanish. Thus, generically, the 
hypersurface $R = {\rm constant}$ does not admit
an integrable complex structure.  The intrinsic torsion analysis can, in principle, be extended to any order
in $\Omega$.

In section \ref{sec:examp} we discuss an example based on an STU-model in the presence of a chiral background
describing the coupling to the square of the Riemann tensor.
Before discussing its intrinsic torsion classes we perform a coordinate change that simplifies the deformed
target-space metric. The coordinate change is the one discussed in \cite{Cardoso:2010gc}.  This coordinate
change is motivated by duality considerations, as follows.  In the presence of a chiral background field,
the usual transformation laws of the fields $Y^I$
under S- and T-duality get modified by terms that depend on the chiral background. It is convenient to 
introduce 
new fields $\tilde{Y}^I$ that under dualities transform in the usual way, i.e. when chiral background fields are absent. These new fields can be defined as follows
\cite{Cardoso:2010gc},
\begin{eqnarray}
{\rm Re} \, \tilde{Y}^I &=& {\rm Re} \, Y^I \;, \nonumber\\
{\rm Re} \, F_I^{(0)} ( \tilde{Y}) &=& {\rm Re} \, F_I ( Y, \bar Y, \Upsilon, \bar \Upsilon) \;.
\label{change-coord-Y}
\end{eqnarray}
The duality transformations for the fields $\tilde{Y}^I$ are independent of $\Upsilon$ and $\bar \Upsilon$,
whereas those of $Y^I$ depend on the chiral background.  Applying the change of coordinates 
\eqref{change-coord-Y} to the STU-model in question and 
performing the intrinsic torsion analysis, we find that at first order in the deformation parameter $\Upsilon$
the target manifold is almost-K\"ahler, with non-vanishing
intrinsic torsion ${\cal W}_2$ (and ${\cal W}_5)$.

In section \ref{sec:hessep} we return to the deformed sigma-model \eqref{lag-sig-def} in terms of the fields
$Y^I$ and $\Upsilon$.
We introduce real coordinates and express the deformed sigma-model in terms of derivatives of a Hesse potential $H$
that is obtained by Legendre transformation of the Lagrangian ${\cal L} = {\rm Im} \, F - \Omega$.  As shown in \cite{Cardoso:2012nh}, the combination ${\cal L}$ arises naturally in the context of point-particle Lagrangians 
that depend on coordinates and velocities.  As mentioned above, the Lagrangian of such a system can be reformulated in terms of
complex coordinates, and in these coordinates it equals ${\cal L}$.  The Hesse potential $H$ is then the
Hamiltonian of this system.  In the absence of a chiral background, the expressions we get reduce to those
obtained recently in \cite{Mohaupt:2011aa}. We conclude with a remark.

\section{The sigma-model in the presence of a chiral background field \label{sec:defsm}}

We consider the scalar field sigma-model that arises in four-dimensional $N=2$ supergravity theories 
based on abelian vector multiplets coupled to supergravity in the presence of an arbitrary chiral background field
\cite{deWit:1996ix,deWit:1996ag}. These theories can be conveniently
described in terms of the superconformal multiplet calculus 
\cite{deWit:1979ug,deWit:1980tn,deWit:1984pk,deWit:1984px}.  At the Wilsonian level, the sigma-model
is encoded in a holomorphic function $F$ that is homogeneous of degree two.  This function gets extended to the non-holomorphic function \eqref{eq:F-0-Om}
when
deforming special geometry by non-holomorphic terms \cite{LopesCardoso:2006bg,Cardoso:2008fr,Cardoso:2010gc}
(see \cite{Cardoso:2012nh} for a review and applications).
In this section, we discuss
the implications of this modification for the sigma-model.

We begin with an analysis of the sigma-model 
in the presence of a chiral
background field at the Wilsonian level. Subsequently we extend the discussion and include
the non-holomorphic deformation of 
special geometry.

\subsection{The sigma-model Lagrangian at the Wilsonian level}

We consider the Wilsonian action describing the coupling of $n$ abelian vector fields 
to supergravity in the presence of an arbitrary chiral background field. This action can be constructed
in a transparent way by making use of the superconformal multiplet calculus, which incorporates
the gauge symmetries of the $N=2$ superconformal algebra. To obtain an action that is gauge equivalent
to a Poincar\'e supergravity theory, two compensating multiplets
need to be coupled to conformal supergravity.  One is a hyper multiplet, that will be omitted in the following, 
since it will not enter in the discussion of the sigma-model.  The other is an abelian vector multiplet.
We will therefore consider $n + 1$ abelian vector multiplets that 
will be labelled by an index 
$I=0, 1, \dots, n$, with $I=0$ referring to the compensating multiplet.
Each of them contains a complex scalar field $X^I$ with Weyl weight 
 $w=1$ and chiral weight $c=-1$.  We will allow for the presence of an arbitrary chiral background superfield,
 whose lowest component field is a bosonic field $\hat A$ that is complex and has Weyl weight $w$ and chiral weight 
 $c = -w$.  
 
The coupling of the vector multiplets to conformal supergravity is encoded
in a holomorphic function $F$ that is homogeneous of degree two.  In the presence of the chiral background, $F$ depends
on the complex scalar fields $X^I$ as well as on $\hat A$, so that the homogeneity condition takes the form
\begin{equation}
F(\lambda \, X,\lambda^w \, \hat{A})=\lambda^2 \, F(X,\hat{A}) \;,
\label{homogeneity-wils}
\end{equation}
where $\lambda \in \mathbb{C} \backslash \{ 0 \}$. 
Therefore the function $F(X, \hat A)$ satisfies the relation
\begin{equation}
2 F = X^I F_I + w \, \hat A F_{\hat A} \;,
\end{equation}
where $F_I$ and $F_{\hat A}$ denote the derivatives of $F(X,\hat{A})$ with respect to $X^I$ and $\hat A$,
respectively.  

As is well known, the field equations of the vector multiplets are subject to equivalence transformations corresponding to electric/magnetic duality transformations. They act as 
${\rm Sp} (2n+2; \mathbb{R})$ linear transformations  on the $(2n+2)$-component
vector $(X^I, F_I(X, \hat A))$.  While the background field $\hat A$ is inert under duality transformations, 
it nevertheless enters in the explicit form of the transformations.

We introduce the following expressions \cite{LopesCardoso:2000qm},
\begin{eqnarray}
{\rm e}^{- \cal K} &=& i \left( {\bar X}^{I} \, F_I(X, \hat A)  - {\bar F}_{\bar I} (\bar X, 
\bar{\hat A}) \, X^I  \right) \;, \nonumber\\
{\cal A}_{\mu} &=&  \tfrac12 \, {\rm e}^{\cal K} \left( \bar{X}^{I} \, 
\overleftrightarrow{{\cal D}_{\mu}} 
\, F_I - 
{\bar F}_{\bar I} \, \overleftrightarrow{{\cal D}_{\mu}} 
 \, X^I \right)  \;,
\label{express-kappa-cala}
\end{eqnarray}
as well as
\begin{equation}
N_{IJ} = - i \left(F_{IJ} - {\bar F}_{\bar I \bar J}\right) \;,
\end{equation}
where
$F_{IJ} =  \partial^2 F(X, \hat A)/\partial X^I \partial X^J$.  Observe that ${\rm e}^{- \cal K}$ has
weights $w=2, c=0$, while $N_{IJ}$ has weights $w = c = 0$.  The covariant derivative ${\cal D}_{\mu}$ is
covariant with respect to chiral $U(1)$ transformations and dilations, and the associated gauge connections
are $A_{\mu}$ and $b_{\mu}$, respectively.  Note that
$\overleftrightarrow{{\cal D}_{\mu}} = {\cal D}_{\mu} - \overleftarrow{{\cal D}_{\mu}} $.
Using \eqref{express-kappa-cala}, we infer the following expression 
for the $U(1)$ connection $A_{\mu}$,
\begin{equation}
A_{\mu} = - \tfrac12 \, {\rm e}^{\cal K} \left( \bar{X}^{I} \, \overleftrightarrow{\partial_{\mu} }
\, F_I - 
{\bar F}_{\bar I} \, \overleftrightarrow{\partial_{\mu} } \, X^I \right) + {\cal A}_{\mu} \;.
\label{eq:U1-conn}
\end{equation}
The combination ${\cal A}_{\mu}$ is $U(1)$-invariant, and it vanishes in the absence of a chiral background.
In the presence of a chiral background, ${\cal A}_{\mu}$ takes a complicated form that 
involves the quantity $F_{\hat A} = \partial F / \partial
{\hat A}$ and derivatives thereof.  
In the case that the chiral background superfield is taken to be the square of the Weyl superfield, the expression for ${\cal A}_{\mu}$ can be found in \cite{LopesCardoso:2000qm}.

Next, we introduce the rescaled Weyl and $U(1)$ invariant fields \cite{LopesCardoso:2000qm}
\begin{eqnarray}
Y^I &=& {\rm e}^{{\cal K}/2} \, {\bar h} \, X^I \;, \nonumber\\
\Upsilon & =& \left( {\rm e}^{{\cal K}/2} \, {\bar h} \right)^w  {\hat A} \;,
\label{Y-Ups-res}
\end{eqnarray}
where $h$ denotes a phase factor which transforms under $U(1)$ with the same weight as the fields $X^I$.
Using the homogeneity property \eqref{homogeneity-wils}, we obtain $F(Y, \Upsilon) = {\rm e}^{\cal K} \, {\bar h}^2 \,
F(X, \hat A)$ as well as
\begin{equation}
i \left( {\bar Y}^{I} \, F_I(Y, \Upsilon)  - {\bar F}_{\bar I} (\bar Y, 
\bar{\Upsilon}) \, Y^I  \right) = 1 \;, 
\label{Y0-rel}
\end{equation}
where $F_I(Y, \Upsilon) = \partial F(Y, \Upsilon) / \partial Y^I$ (and similarly $F_{\Upsilon} = 
\partial F(Y, \Upsilon) / \partial \Upsilon$).  Expressing \eqref{eq:U1-conn} in terms of 
the rescaled fields \eqref{Y-Ups-res} yields the $U(1)$-invariant combination
\begin{equation}
a_{\mu} \equiv
A_{\mu} - i \partial_{\mu} \ln h 
= - \tfrac12 \, \left( \bar{Y}^{I} \, \overleftrightarrow{\partial_{\mu} }
\, F_I(Y, \Upsilon)  - 
{\bar F}_{\bar I}(\bar Y, \bar \Upsilon)  \, \overleftrightarrow{\partial_{\mu} }\, Y^I \right) + {\cal A}_{\mu} \;.
\label{eq:U1-conn-res}
\end{equation}
We express ${\cal D}_{\mu} X^I$ in terms of the rescaled fields \eqref{Y-Ups-res}.
We use the invariance under special conformal transformations to set $b_{\mu} =0$, and we obtain
\begin{equation}
{\rm e}^{{\cal K}/2} \, {\bar h} \, {\cal D}_{\mu} X^I = \partial_{\mu} Y^I + i a_{\mu} \, Y^I \;.
\end{equation}

Now we turn to the kinetic term for the scalar fields $X^I$ in the Wilsonian Lagrangian,
\begin{equation}
L = i \left( {\cal D}^{\mu} F_I(X,\hat A) \, {\cal D}_{\mu} {\bar X}^{I} - 
{\cal D}^{\mu} {\bar F}_{\bar I}(\bar X,\bar{\hat A}) \, {\cal D}_{\mu} X^I \right) \;,
\end{equation}
which we rewrite in terms of the rescaled fields \eqref{Y-Ups-res}, 
\begin{equation}
L = {\rm e}^{- \cal K} \left[ i \left( \partial^{\mu} F_I(Y,\Upsilon) \, \partial_{\mu} {\bar Y}^{I} - 
\partial^{\mu} {\bar F}_{\bar I}(\bar Y,\bar{\Upsilon}) \, \partial_{\mu} Y^I \right) 
- \left( a^{\mu} - {\cal A}^{\mu} \right)  
\left(a_{\mu} - {\cal A}_{\mu} \right) + {\cal A}^{\mu} {\cal A}_{\mu} \right]\;.
\label{lag-interm}
\end{equation}
Observe that the combination $a_{\mu} - {\cal A}_{\mu}$ and ${\cal A}_{\mu}$ do not couple to one another.
The combination $a_{\mu} - {\cal A}_{\mu}$, which is given in \eqref{eq:U1-conn-res},
is determined in terms of $(Y^I, F_I(Y, \Upsilon))$, and is non-vanishing in the absence of a chiral background.
The term ${\cal A}^{\mu} {\cal A}_{\mu}$, on the other hand, vanishes when switching off the chiral background.
When the chiral background is identified with the square of the Weyl tensor, this term has a structure that is
distinct from those of the other terms in \eqref{lag-interm}.
For instance, ${\cal A}^{\mu} {\cal A}_{\mu}$ contains a term proportional
to $T^+ T^- T^+ T^-$, where $T^{\pm}$ refers to (anti)self-dual Lorentz tensors.  
The other terms in \eqref{lag-interm} do not contain such a term, since they
are constructed out of derivatives of the
vector $(Y^I, F_I(Y, \Upsilon))$.
Also note that ${\cal A}^{\mu} {\cal A}_{\mu}$ is of order ${\cal O}(F_{\Upsilon}^2)$
and higher, and can thus be neglected when working to first order in $F_{\Upsilon}$. Thus, in the following,
we split the Wilsonian Lagrangian into two pieces, $L_{\sigma}^{\rm Wils} + L_{\rm grav}$, where 
$L_{\sigma}^{\rm Wils}$ denotes
the sigma-model Lagrangian that we will be focussing on in the following,
\begin{equation}
L_{\rm \sigma}^{\rm Wils} = {\rm e}^{- \cal K} \left[ i \left( \partial^{\mu} F_I(Y,\Upsilon) \, \partial_{\mu} {\bar Y}^{I} - 
\partial^{\mu} {\bar F}_{\bar I}(\bar Y,\bar{\Upsilon}) \, \partial_{\mu} Y^I \right) 
- 
\left( a^{\mu} - {\cal A}^{\mu} \right)  \left(a_{\mu} - {\cal A}_{\mu} \right) 
\right]\;,
\label{lag-sigm}
\end{equation}
while $L_{\rm grav}$ denotes the remaining part of the supergravity Lagrangian, which includes the term
${\cal A}^{\mu} {\cal A}_{\mu}$.  The overall factor $ {\rm e}^{- \cal K}$ appearing in 
\eqref{lag-sigm} may be absorbed by a rescaling of the space-time metric or, equivalently, set to a constant value
by using the freedom under dilations. In the following, we will thus consider \eqref{lag-sigm}
with ${\rm e}^{- \cal K} =1$.

The sigma-model Lagrangian \eqref{lag-sigm}, which is entirely constructed out of derivatives of
the vector $(Y^I, F_I(Y, \Upsilon))$, is manifestly invariant
under symplectic transformations of $(Y^I, F_I(Y, \Upsilon))$, which leave $\Upsilon$ inert.
We now deform this Wilsonian sigma-model Lagrangian by allowing for non-holomorphic deformations
that are incorporated into $F$ by performing the extension \eqref{eq:F-0-Om}.
The function $F(Y, \bar Y, \Upsilon, \bar{\Upsilon})$, which is 
homogeneous of degree two, can be decomposed into a holomorphic background independent piece $F^{(0)}(Y)$
and a real function $\Omega(Y, \bar Y, \Upsilon, \bar{\Upsilon})$ that encodes the background dependence.
This decomposition is defined up to an anti-holomorphic function, and the associated equivalence
transformation takes the form \cite{Cardoso:2012nh}
\begin{equation}
F^{(0)}(Y) \rightarrow F^{(0)}(Y) + g(Y, \Upsilon) \;\;\;,\;\;\;\; \Omega \rightarrow \Omega - {\rm Im} \, g(Y, \Upsilon) \;,
\label{equival-transf}
\end{equation}
resulting in $F(Y, \bar Y, \Upsilon, \bar{\Upsilon})  \longrightarrow F(Y, \bar Y, \Upsilon, \bar{\Upsilon})
 + {\bar g}(\bar Y, \bar \Upsilon)$.
The latter leaves the vector $(Y^I,$ $F_I(Y, \bar Y, \Upsilon, \bar \Upsilon))$ unaffected.
Electric/magnetic duality now acts as an ${\rm Sp} (2n+2, \mathbb{R})$- transformation on  this vector.

Thus, in the presence of non-holomorphic terms, the deformed sigma-model Lagrangian will be specified by
\begin{equation}
L_{\rm \sigma} =  i \left( \partial^{\mu} F_I(Y,\bar{Y}, \Upsilon, \bar{\Upsilon}) \, \partial_{\mu} {\bar Y}^{I} - 
\partial^{\mu} {\bar F}_{\bar I}(Y, \bar Y, \Upsilon, \bar{\Upsilon}) \, \partial_{\mu} Y^I \right) 
- \mathscr{A}^{\mu} \mathscr{A}_{\mu}
\;,
\label{lag-sig-def}
\end{equation}
where
\begin{equation}
\mathscr{A}_{\mu} = 
- \tfrac12 \, \left( \bar{Y}^{I} \, \overleftrightarrow{\partial_{\mu} }
\, F_I(Y, \bar Y,  \Upsilon, \bar \Upsilon)  - 
{\bar F}_{\bar I}(Y, \bar Y, \Upsilon, \bar \Upsilon)  \, \overleftrightarrow{\partial_{\mu} }\, Y^I \right) \;,
\label{A-conn-def}
\end{equation}
with the $Y^I$ satisfying the relation
\begin{equation}
i \left( {\bar Y}^{I} \, F_I(Y, \bar Y, \Upsilon, \bar \Upsilon)  - {\bar F}_{\bar I} (Y, \bar Y, \Upsilon,
\bar{\Upsilon}) \, Y^I  \right) = 1 \;.
\label{Y0-rel-def}
\end{equation}
The Wilsonian limit is recovered by taking $\Omega$ to be harmonic,
\begin{equation}
\Omega_{\rm Wils} (Y, \bar Y, \Upsilon, \bar \Upsilon) = f(Y, \Upsilon) +  \bar{f}(\bar{Y}, \bar{\Upsilon}) \;,
\label{wilson-om}
\end{equation}
which, through the equivalence transformation \eqref{equival-transf}, results in $F(Y, \Upsilon) = F^{(0)}(Y) + 2 i 
 f(Y, \Upsilon)$.  In the next section, we turn to the evaluation of the deformed sigma-model Lagrangian
\eqref{lag-sig-def}.

\subsection{The sigma-model Lagrangian in the presence of non-holomorphic terms}

The deformed sigma-model $L_{\sigma}$ introduced above is defined in terms of fields $(Y^I, \Upsilon)$.  In the following, 
we express this sigma-model in terms of projective coordinates given by
\begin{eqnarray}
z^i &=&  \frac{Y^i}{Y^0} \;\;\;,\;\;\; i = 1, \dots, n \;, \nonumber\\
\Psi &=&  \frac{\Upsilon}{\left(Y^0\right)^w} \;, \nonumber\\
R &=&  \frac{\bar{Y}^0}{Y^0} \;.
\label{project-coord}
\end{eqnarray}
Here $R = 1/{\bar R}$ denotes the phase of $(Y^0)^{-2}$.  Observe that all these coordinates (including $\Psi$)
transform under symplectic transformations of $(Y^I, F_I)$.
The norm of $Y^0$ is expressed in terms of these
projective coordinates using the relation \eqref{Y0-rel-def}, as follows.  Using the homogeneity of $F(Y, \bar Y, \Upsilon, \bar \Upsilon)$, we obtain
\begin{equation}
F(Y, \bar Y, \Upsilon, \bar \Upsilon) = \left( Y^0 \right)^2 \, {\cal F}(z, {\bar z}, \Psi, \bar \Psi, R) 
\end{equation}
with
\begin{eqnarray}
{\cal F}(z, {\bar z}, \Psi, \bar \Psi, R) &=& {\cal F}^{(0)}(z) + 2i \, \Omega(z, {\bar z}, \Psi, \bar \Psi, R) \;,
\nonumber\\
{\bar {\cal F}}(z, {\bar z}, \Psi, \bar \Psi, \bar{R}) &=& \bar{{\cal F}}^{(0)}(\bar{z}) - 2i \, \Omega(\bar{z}, z, 
\bar{\Psi}, \Psi, \bar{R}) \;,
\label{cal-F-om}
\end{eqnarray}
where $\Omega(z, {\bar z}, \Psi, \bar \Psi, R)$ is not any longer a real function due to its dependence on the phase
$R$.  Using the expressions for the 
first-order derivatives of $F(Y, \bar Y, \Upsilon, \bar \Upsilon)$ given in \eqref{der1-F} and inserting 
these expressions into 
\eqref{Y0-rel-def} yields
\begin{equation}
|Y^0|^2 = - i \, \Sigma^{-1}
\label{norm-Y0-sig}
\end{equation}
with $\Sigma$ given by
\begin{equation}
\Sigma(z, \bar{z}, \Psi, \bar{\Psi}, R) = 2 \left( {\cal F} - {\bar {\cal F}} \right) - \left(z^i - {\bar z}^i \right) \left( {\cal F}_i + 
{\bar {\cal F}}_{\bar{\imath}} \right) - w \left( \Psi {\cal F}_{\Psi} - \bar{\Psi} \bar{\cal F}_{\bar{\Psi} }\right)
- \left( R \,  {\cal F}_{R} - \bar{R} \, \bar{\cal F}_{\bar{R}}\right) \;.
\label{eq:Sigma-comb-psi}
\end{equation}
Observe that $\Sigma$ satisfies $\Sigma = - \bar \Sigma$, and that it can be expressed as 
$\Sigma = \Delta - {\bar \Delta}$ with
\begin{equation}
\Delta(z, \bar{z}, \Psi, \bar{\Psi}, R) = 2 \, {\cal F} - \left(z^i - {\bar z}^i \right) {\cal F}_i  - w \,\Psi {\cal F}_{\Psi} 
-  R \,  {\cal F}_{R} \;.
\label{delta-z-ps}
\end{equation}

Now we express $\mathscr{A}^{\mu} $ given in \eqref{A-conn-def} in terms of the coordinates
\eqref{project-coord}.  First, using \eqref{Y0-rel-def}, we note that $\mathscr{A}_{\mu}$ can be written as
\begin{equation}
 \mathscr{A}_{\mu}  =
  \bar F_{\bar I}\partial_{\mu}Y^I-\bar Y^I\partial_{\mu}F_I  = F_I \partial_{\mu} \bar{Y}^{I} - Y^I \partial_{\mu}
  \bar{F}_{\bar I} \;.
  \end{equation}
Subsequently we obtain  
\begin{eqnarray}
\label{A-a-conn}
 \mathscr{A}_{\mu}
 &=& - \Sigma \, \bar{Y}^0 \, \partial_{\mu} Y^0 - |Y^0|^2 \, \mathfrak{a}_{\mu}=
 - \bar{\Sigma} \, {Y}^0 \, \partial_{\mu} {\bar Y}^0 - |Y^0|^2 \, \bar{\mathfrak{a}}_{\mu} \;, \nonumber\\
\mathfrak{a}_{\mu} &=& - \bar{\cal F}_{\bar \imath} \partial_{\mu}
z^i - {\cal F}_i \partial_{\mu} {\bar z}^i + \partial_{\mu} \Delta  \\
&=&
\left( \Delta_i - \bar{\cal F}_{\bar \imath} \right)
\partial_{\mu} z^i + 
\left( \Delta_{\bar \imath} - {\cal F}_i \right)
\partial_{\mu} {\bar z}^i
+ \Delta_{\Psi} \partial_{\mu} \Psi + \Delta_{\bar \Psi} \partial_{\mu} {\bar \Psi} + \Delta_R \partial_{\mu} R 
\;, \nonumber
\end{eqnarray}
as well as 
\begin{eqnarray}
\mathscr{A}^{\mu} \mathscr{A}_{\mu}
&=& |\Sigma|^2 \, |Y^0|^2 \, \partial_{\mu} Y^0 \partial^{\mu} {\bar Y}^0
+ |Y^0|^4 \, \mathfrak{a}_{\mu} \bar{\mathfrak{a}}^{\mu} \;, \nonumber\\
&& + \Sigma \, |Y^0|^2 \, {\bar Y}^0 \partial_{\mu} Y^0 \, \bar{\mathfrak{a}}^{\mu}
+ \bar \Sigma \, |Y^0|^2 \, Y^0 \, \partial_{\mu} {\bar Y}^0 \, \mathfrak{a}^{\mu} \;.
\end{eqnarray}
The field strength of $\mathscr{A}_{\mu}$ reads
\begin{eqnarray}
\mathscr{F}_{\mu \nu} &=& \partial_{\mu} \mathscr{A}_{\nu} - \partial_{\nu} \mathscr{A}_{\mu} \nonumber\\
&=& 
\partial_{\mu} F_I \partial_{\nu} \bar{Y}^{I} - \partial_{\mu} Y^I \partial_{\nu}
  \bar{F}_{\bar I} - ( \mu \leftrightarrow \nu) \nonumber\\
&=& \frac{i}{\Sigma} \left[ \partial_{\mu} \, \mathfrak{a}_{\nu} - \partial_{\nu} \mathfrak{a}_{\mu} - 
\frac{1}{\Sigma} \left( \partial_{\mu} \Sigma \, \mathfrak{a}_{\nu} - 
\partial_{\nu} \Sigma \, \mathfrak{a}_{\mu} \right) \right] \;,
\label{field-A}
\end{eqnarray}
where we used \eqref{norm-Y0-sig}.

We proceed to express the sigma-model Lagrangian \eqref{lag-sig-def} in terms of projective coordinates.  
Using the expressions for the second-order derivatives of $F$ given in \eqref{der2-F} we obtain,
\begin{eqnarray}
F_{00} + F_{0i} (z^i + \bar z^i) + F_{ij} \, z^i {\bar z}^j + w \Psi \left(Y^0\right)^{w-1} \, \left(F_{0 \Upsilon}
+ F_{i \Upsilon} \, {\bar z}^i \right) &=& \Delta - R \, \Delta_R \;, \nonumber\\
F_{0i} - {\bar F}_{\bar 0 \bar \imath} + \left(F_{ij} - {\bar F}_{\bar \imath \bar \jmath} \right) z^j
+ F_{i \Upsilon} \, w \Psi\, \left(Y^0\right)^{w-1} &=& {\cal F}_i - {\bar \Delta}_{\bar \imath} - R \, {\cal F}_{i R }
\;, \nonumber\\
\left( F_{0 \Upsilon} + F_{i \Upsilon} \, {\bar z}^i \right) \left( Y^0 \right)^{w-1} &=& \Delta_{\Psi} \;,
\nonumber\\
F_{0 \bar 0} + F_{0 \bar \imath} {\bar z}^{i} + F_{i \bar 0} {\bar z}^i + F_{i \bar \jmath} {\bar z}^i {\bar z}^j
+ F_{0 \bar \Upsilon} w {\bar \Psi} \left( {\bar Y}^0 \right)^{w-1} + F_{i \bar \Upsilon} {\bar z}^i 
w {\bar \Psi}  \left( \bar Y^0 \right)^{w-1} &=& \Delta_R \;, \nonumber\\
F_{0 \bar \imath} + F_{i \bar 0}  + \left( F_{j \bar \imath}  +  F_{i \bar \jmath} \right) {\bar z}^j 
+ F_{i \bar \Upsilon} w {\bar \Psi} \left( {\bar Y}^0 \right)^{w-1} &=& {\cal F}_{i R}
+ {\bar R} \left( \Delta_{\bar \imath} - {\cal F}_i \right) \;, \nonumber\\
\left( F_{0 \bar \Upsilon} + F_{i \bar \Upsilon} {\bar z}^i \right) \left( {\bar Y}^0 \right)^{w-1} &=& \Delta_{\bar \Psi} \;. \nonumber\\
\end{eqnarray}
Using these we get
\begin{eqnarray}
i\left(\partial_{\mu}F_I\partial^{\mu}\bar Y^I-\partial_{\mu} \bar F_{\bar I}\partial^\mu Y^I\right)
&=& i \Big[ \partial_{\mu} Y^0 \partial^{\mu} {\bar Y}^0 \left( \Sigma - R \Delta_R + {\bar R} \Delta_{\bar R} \right)
\nonumber\\
&& \quad + \partial_{\mu} Y^0 \partial^{\mu} {\bar z}^i \, {\bar Y}^0 \left( {\cal F}_i - {\bar \Delta}_{\bar \imath} - R {\cal F}_{i R }
\right) \nonumber\\
&& \quad 
- \partial_{\mu} {\bar Y}^0 \partial^{\mu} { z}^i \, {Y}^0 \left( {\bar {\cal F}}_{\bar \imath} - {\Delta}_i - 
{\bar R} \bar{\cal F}_{\bar \imath \bar R }
\right) \nonumber\\
&& \quad + \partial_{\mu} z^i \partial^{\mu} {\bar z}^j \, |Y^0|^2 \left( {\cal F}_{ij} - {\bar {\cal F}}_{\bar \imath \bar \jmath} \right) \nonumber\\
&& \quad + \partial_{\mu} \Psi \partial^{\mu} {\bar Y}^0 \, Y^0 \, \Delta_{\Psi} 
- \partial_{\mu} {\bar \Psi} \partial^{\mu} {Y}^0 \, {\bar Y}^0 \, {\bar \Delta}_{\bar \Psi} \nonumber\\
&& \quad  + \partial_{\mu} \Psi \partial^{\mu} {\bar z}^i \, |Y^0|^2 \, {\cal F}_{i \Psi} - 
 \partial_{\mu} {\bar \Psi} \partial^{\mu} {z}^i \, |Y^0|^2 \, {\bar {\cal F}}_{\bar \imath \bar \Psi}
\Big] \nonumber\\
&& + i \Big[ \partial_{\mu} {\bar Y}^0 \partial^{\mu} {\bar Y}^0 \, \Delta_R  \nonumber\\
&& \quad + \partial_{\mu} {\bar Y}^0 \partial^{\mu} {\bar z}^i \, {\bar Y}^0 
\left( {\cal F}_{i R}  + \bar R \left(  {\Delta}_{\bar \imath} - 
{\cal F}_i
\right) \right) \nonumber\\
&& \quad + \partial_{\mu} {\bar z}^i \partial^{\mu} {\bar z}^j \, |Y^0|^2 \, {\cal F}_{i \bar \jmath} \nonumber\\
&& \quad + \partial_{\mu} {\bar Y}^0 \partial^{\mu} \bar \Psi \, Y^0 \, \Delta_{\bar \Psi} \nonumber\\
&& \quad + \partial_{\mu} {\bar z}^i \partial^{\mu} \bar \Psi \, |Y^0|^2 \, {\cal F}_{i \bar \Psi} - {\rm c.c.} \Big] \;. 
\label{rel_2der-F}
\end{eqnarray}
Next, using the relation
\begin{equation}
\partial_{\mu} R = \frac{\partial_{\mu} \bar Y^0}{Y^0} - R \, \frac{\partial_{\mu} Y^0}{Y^0} \;,
\end{equation}
as well as \eqref{norm-Y0-sig} with $\Sigma = - \bar \Sigma$, we obtain
\begin{eqnarray}
 L_{\sigma}&=& i |Y^0|^2 \Big[\partial_{\mu} z^i \partial^{\mu} {\bar z}^j 
\left( {\cal F}_{ij} - {\bar {\cal F}}_{\bar \imath \bar \jmath} \right) \nonumber\\
&& \qquad \quad + \partial_{\mu} \Psi \partial^{\mu} {\bar z}^i \, {\cal F}_{i \Psi} 
- \partial_{\mu} \bar \Psi \partial^{\mu} {z}^i \, \bar{\cal F}_{\bar \imath \bar \Psi} 
\nonumber\\
&& \qquad \quad + \partial_{\mu} R \partial^{\mu} {\bar z}^i \, {\cal F}_{i R} 
- \partial_{\mu} \bar R \partial^{\mu} {z}^i \, \bar{\cal F}_{\bar \imath \bar R} 
\nonumber\\
&& \qquad \quad + \Big( \partial_{\mu} \bar z^i \partial^{\mu} {\bar z}^j 
{\cal F}_{i \bar \jmath} + \partial_{\mu} \bar z^i \partial^{\mu} \bar \Psi
{\cal F}_{i \bar \Psi} - {\rm c.c.} \Big) \Big] - |Y^0|^4 \, \mathfrak{a}_{\mu} \bar{\mathfrak{a}}^{\mu} \;.
\label{lag-def-proj}
\end{eqnarray} 
Introducing $\rho^M = (z^i, \bar z^i, \Psi, \bar \Psi, R)$, 
this can also be written as
\begin{eqnarray}
 L_{\sigma}= 
 i |Y^0|^2 \Big[\partial_{\mu} \rho^M  \partial^{\mu} {\bar z}^j \,
{\cal F}_{Mj} - {\rm c.c.} \Big] - |Y^0|^4 \, \mathfrak{a}_{\mu} \bar{\mathfrak{a}}^{\mu} \;,
\label{eq:lag-nonhol-final}
\end{eqnarray} 
with $\mathfrak{a}_{\mu}$ given in \eqref{A-a-conn}.
Thus we find that the deformed sigma-model Lagrangian takes a rather simple form,
with the 
first term only depending on double derivatives of ${\cal F}(z, \bar z,\Psi, \bar \Psi, R)$.
The Lagrangian \eqref{eq:lag-nonhol-final} describes the coupling of the deformed
sigma-model to supergravity.  In the rigid limit, i.e. when decoupling supergravity
in which case $\mathfrak{a}_{\mu} =0$, the deformed sigma-model Lagrangian
yields a generalization of the extrinsic construction of special-K\"ahler manifolds given in \cite{Alekseevsky:1999ts}.
We will turn to this construction in the last subsection.

Note also that the Lagrangian \eqref{eq:lag-nonhol-final}
depends on the phase of $Y^0$, i.e. on $R$. This means that when coupling the deformed
sigma-model to supergravity,  
the associated
target-space line element  will depend on this coordinate, in general.  This dependence
drops out in the Wilsonian limit when taking $(z^i, \Psi)$ to be independent fields, as we will now show.

\subsubsection{The Wilsonian limit \label{sec:wils}}

In the Wilsonian limit \eqref{wilson-om}, we have
\begin{eqnarray}
F_{\rm Wils}(Y, \bar Y, \Upsilon, \bar \Upsilon) &=& \left( Y^0 \right)^2 \, {\cal F}_{\rm Wils} (z, {\bar z}, \Psi, \bar \Psi, R) \;, \nonumber\\
{\cal F}_{\rm Wils}(z, {\bar z}, \Psi, \bar \Psi, R) &=& {\cal F}^H (z, \Psi)
+ 2 i \, R^2 \, \bar{f} (\bar{z}, \bar{\Psi})  \;, 
\label{eq:calF-wils}
\end{eqnarray}
where ${\cal F}^H$ denotes the holomorphic part of ${\cal F}_{\rm Wils}$, 
\begin{equation}
{\cal F}^H (z, \Psi) = 
{\cal F}^{(0)}(z) + 2 i \, f(z, \Psi) \;.
\end{equation}
Inserting \eqref{eq:calF-wils} into \eqref{eq:Sigma-comb-psi} results in
\begin{equation}
\Sigma^{\rm Wils} =  2 \left( {\cal F}^H - \bar{\cal F}^H \right) - 
\left(z^i - {\bar z}^i \right) \left( {\cal F}_i^H +
{\bar {\cal F}}_{\bar{\imath}}^H \right) -  w \left( \Psi \, {\cal F}_{\Psi}^H - \bar{\Psi} \, \bar{{\cal F}}_{\bar{\Psi} }^H \right) \;.
\label{eq:Sigma-comb-wils}
\end{equation}
Note that, as expected, the 
dependence on $R$ and $\bar R$ has dropped out of $\Sigma^{\rm Wils}$. As mentioned below \eqref{wilson-om},
$F_{\rm Wils}$ and $(Y^0)^2 {\cal F}^H (z, \Psi)$ give rise to equivalent Wilsonian Lagrangians, and thus 
$\Sigma^{\rm Wils}$ can only depend on ${\cal F}^H$ and derivatives thereof.

Using the expressions \eqref{der1-delta} we find
\begin{eqnarray}
\Delta_R & = & 0 \;, \nonumber\\
\Delta_{\Psi} &=& \Sigma_{\Psi}^{\rm Wils} \;, \nonumber\\
\Delta_{\bar \Psi} &=& 0 \;, \nonumber\\
\Delta_{\bar \imath} - {\cal F}_i &=& 0 \;, \nonumber\\
{\cal F}_i - {\bar \Delta}_{\bar \imath} &=& \Sigma^{\rm Wils}_{\bar \imath} \;, \nonumber\\
{\cal F}_{i \Psi} &=& \Sigma^{\rm Wils}_{\bar \imath \Psi} \;, \nonumber\\
{\cal F}_{ij} - {\bar{\cal F}}_{\bar \imath \bar \jmath} &=& \Sigma^{\rm Wils}_{i \bar \jmath} \;,
\nonumber\\
{\cal F}_{iR} &=& 0 \;, \nonumber\\
\Sigma_{\Psi \bar \Psi}^{\rm Wils} &=& 0 \;.
\label{eq:wils-rel}
\end{eqnarray}
Inserting \eqref{eq:wils-rel} into \eqref{A-a-conn} and \eqref{eq:lag-nonhol-final}, we obtain
\begin{eqnarray}
\mathfrak{a}_{\mu} =  \Sigma^{\rm Wils}_i \, \partial_{\mu} z^i + \Sigma^{\rm Wils}_{\Psi} \, \partial_{\mu} \Psi \;,
\label{A-wils}
\end{eqnarray}
as well as 
\begin{eqnarray}
 L_{\sigma}^{\rm Wils} = |Y^0|^2 \Big[ i \Sigma_{A \bar B} \, \partial_{\mu} v^A
 \partial^{\mu} {\bar v}^B - |Y^0|^2 \, \Sigma_A \, \partial_{\mu} v^A \, {\bar \Sigma}_{\bar B} \, 
\partial^{\mu} {\bar v}^B \Big] \;,
\end{eqnarray}
where we introduced $v^A = (z^i, \Psi)$.
Recalling \eqref{norm-Y0-sig} and defining
\begin{equation}
{\rm e}^{-K} = i \Sigma^{\rm Wils} \;,
\end{equation}
we obtain
\begin{eqnarray}
 L_{\sigma}^{\rm Wils}  = - \partial_A \partial_{\bar B} K \, 
 \partial_{\mu} v^A \, 
\partial^{\mu} {\bar v}^B  \;.
\label{sigma-enlar}
\end{eqnarray}
Thus, in the Wilsonian limit, the enlarged target-space 
is K\"ahler, 
with the real K\"ahler potential $K$ determined in terms of ${\cal F}^H (z, \Psi)$
via \eqref{eq:Sigma-comb-wils}. Using \eqref{A-wils} we obtain for the 
field strength 1-form \eqref{field-A}, 
\begin{eqnarray}
\mathscr{F} = i \partial_A \partial_{\bar B} K \, d v^A \wedge d \bar{v}^B \;.
\label{field-A-wils}
\end{eqnarray}

So far, we treated the $v^A$ as independent fields.  Now let us discuss the case when $\Psi$ becomes 
a dependent field.  Let us first consider the restriction to a hypersurface $\Psi = g(z)$.
This is achieved by taking $\Upsilon$ to be the field dependent function $\Upsilon = (Y^0)^w \, g(z)$.
Then, the sigma-model Lagrangian \eqref{sigma-enlar} and the field strength \eqref{field-A-wils}
retain their form,
\begin{eqnarray}
 L_{\sigma}^{\rm Wils}  &=& - \partial_i \partial_{\bar \jmath} K(z, \bar z, \Psi(z), \bar{\Psi}(\bar z) )\, 
 \partial_{\mu} z^i \, 
\partial^{\mu} {\bar z}^{j}  \;, \nonumber\\
\mathscr{F} &=& i \partial_i \partial_{\bar \jmath} K(z, \bar z, \Psi(z), \bar{\Psi}(\bar z) )
 \, d z^i \wedge d \bar{z}^j \;.
 \label{sigma-wil-hyp}
\end{eqnarray}

Next, let us consider the case when $\Upsilon$ is taken to be a constant parameter, in which case
$\Psi = \Upsilon/(Y^0)^w$  
(and similarly $\bar \Psi$)
become functions of $(z, \bar z, R, \Upsilon, \bar \Upsilon)$. 
Namely, using \eqref{norm-Y0-sig}, we obtain
\begin{equation}
\Psi(z, \bar z, R, \Upsilon, \bar \Upsilon) = \Upsilon \left( i \Sigma \, R \right)^{w/2} \;.
\label{psi-sig-R}
\end{equation}
Note that $\Sigma$ depends on $\Psi$ and $\bar \Psi$, so that to obtain $\Psi(z, \bar z, R, \Upsilon, \bar \Upsilon)$
one has to proceed by iteration in $\Upsilon$ and
$\bar \Upsilon$. 
Then, generically, the sigma-model will depend on the coordinates $(z^i, \bar z^i, R)$ and will 
not retain the form \eqref{sigma-wil-hyp}. 
It will have additional terms of the form
$\partial_{\mu} z^i \partial^{\mu} z^j$ as well as terms that involve $\partial_{\mu} R$.

\subsubsection{The rigid case \label{sec:rigid-lim}}

Now we consider the rigid limit of \eqref{eq:lag-nonhol-final}.  The decoupling of 
supergravity proceeds by setting $\mathfrak{a}_{\mu} =0$ and setting $Y^0=1$,
so that
\begin{eqnarray}
 L_{\sigma}^{\rm rigid}= 
 i  \Big[\partial_{\mu} \rho^M  \partial^{\mu} {\bar z}^j \,
{\cal F}_{Mj} - {\rm c.c.} \Big] \;,
\label{eq:lag-nonhol-rigid}
\end{eqnarray} 
where now $\rho^M =  (z^i, \bar z^i, \Psi = \Upsilon, \bar \Psi =\bar \Upsilon)$ and ${\cal F}= {\cal F} (z, \bar{z}, \Psi,
\bar \Psi)$.  This Lagrangian
follows from the rigid Wilsonian Lagrangian
\begin{equation}
i \Big[\partial_{\mu} {\cal F}^H_j  \partial^{\mu} {\bar z}^j - {\rm c.c.} \Big]
\end{equation}
by performing the replacement  ${\cal F}^H_j (z, \Psi) \rightarrow  {\cal F}_j (z, \bar z, \Psi, \bar \Psi)$.  When taking the fields $\Psi$ and $\bar \Psi$ to be constant, i.e. when treating them
as parameters, the Lagrangian \eqref{eq:lag-nonhol-rigid} can be
understood in terms of the extrinsic construction given in  \cite{Alekseevsky:1999ts}, as follows.

We consider the ambient space $V= \mathbb{C}^{2n}$ with local coordinates
$(z^i, w_i)$, standard 
complex symplectic form
\begin{equation}
\Omega_V = dz^i \wedge d w_i \;,
\end{equation}
and standard complex structure
\begin{equation}
J_V = i \left(dz^i \otimes \frac{\partial}{\partial z^i} + dw_i \otimes \frac{\partial}{\partial w_i}
- d{\bar z}^i  \otimes \frac{\partial}{\partial {\bar z}^{i}} - d{\bar w}_{i} \otimes 
\frac{\partial}{\partial {\bar w}_{ i}} 
\right) \;.
\end{equation}
We define the
Hermitian form $\gamma_V$ (i.e. $\gamma_V (J_V\, X, J_V\, Y) = \gamma_V (X, Y) \;
\forall X, Y \, \in {\rm Vec} (V)$),
\begin{equation}
\gamma_V = i \left( dz^i \otimes d{\bar w}_{i} - dw_i \otimes d{\bar z}^{i} \right) \;,
\end{equation}
which yields 
the Riemannian metric  
\begin{equation}
g_V = {\rm Re} \, \gamma_V = i \left( dz^i \otimes_{\rm sym} d{\bar w}_{ i} - dw_i \otimes_{\rm sym} d{\bar z}^{ i} \right) 
\end{equation}
and the fundamental 2-form $\omega_V$ (i.e. $\omega_V (X, Y) = - g_V (J_V\, X, Y)$) 
\begin{equation}
\omega_V = {\rm Im}\, \gamma_V = 
\left( dz^i \wedge d{\bar w}_{i} - dw_i \wedge d{\bar z}^{ i} \right) \;\;\;,\;\;\; d \omega_V =0 \;.
\end{equation}
Now consider the hypersurface $M$ in ambient space, 
described by $w_i = F_{i} (z, \bar z)$ (we supress the dependence on the
parameters $\Psi$ and $\bar \Psi$, for simplicity).  Unlike the case studied in \cite{Alekseevsky:1999ts}, this is
not a holomorphic immersion.
The pullback of
the complex symplectic form $\Omega_V$ gives
\begin{equation}
\Omega_M = F_{i \bar \jmath} \, dz^i \wedge d{\bar z}^{ j} \;,
\end{equation}
which is non-vanishing, and therefore  $w_i = F_{i} (z, \bar z)$ is 
not anylonger a Lagrangian immersion.  The
pullback of the hermitian form $\gamma_V$ gives
\begin{equation}
\gamma_M = N_{ij} \,   dz^i \otimes d{\bar z}^{ j}  + i \, {\bar F}_{\bar \imath j} \, dz^i \otimes dz^j
- i  \, F_{i \bar \jmath} \, d{\bar z}^j \otimes d{\bar z}^{ i}\;,
\end{equation}
where $N$ is a hermitian matrix given by
\begin{equation}
N_{ij} = - i \left( F_{ij} - {\bar F}_{\bar \imath \bar \jmath} \right) \;.
\end{equation}
The induced metric on $M$ is 
\begin{equation}
g_M = {\rm Re} \, \gamma_M = 
N_{ij} \,   dz^i \otimes_{\rm sym} d{\bar z}^{j}  + i \, {\bar F}_{\bar \imath j} \, dz^i \otimes_{\rm sym} dz^j
- i  \, F_{i \bar \jmath} \, d{\bar z}^{ i} \otimes_{\rm sym} d{\bar z}^{j}\;,
\label{indmet}
\end{equation}
while the pullback of the fundamental 2-form $\omega_V$ is
\begin{equation}
\omega_M = {\rm Im} \, \gamma_M = - i \, 
N_{ij} \,   dz^i \wedge d{\bar z}^{j}  + \, {\bar F}_{\bar \imath j} \, dz^i \wedge dz^j
+  \, F_{i \bar \jmath} \, d{\bar z}^{ i} \wedge d{\bar z}^{j} \;\;\;,\;\;\; d \omega_M =0 \;.
\end{equation}
The induced metric $g_M$ is not hermitian with respect to the standard complex structure
\begin{equation}
J_M \left( \frac{\partial}{\partial z^i} \right) = i \, \frac{\partial}{\partial z^i} \;\;\;,\;\;\;
J_M \left( \frac{\partial}{\partial {\bar z}^{i}} \right) = - i \, \frac{\partial}{\partial {\bar z}^{i}} \;,
\end{equation}
i.e. $g_M (J_M \, X, J_M \, Y) \neq g_M (X, Y)$ due to the terms $F_{i \bar \jmath}$.

On the other hand, it is well known that 
given a non-degenerate 2-form $\omega_M$
and a non-degenerate metric $g_M$, it is possible to construct an almost complex structure $J_M$ such that 
\begin{equation}
\omega_M (J_MX,J_MY) = \omega_M( X, Y).
\end{equation}
The construction proceeds as follows. The non-degeneracy of both  $g_M$ and $\omega_M$
implies the existence of a linear map $A$ such that 
\begin{equation}
g_M (A X,Y) = -\omega_M( X, Y).
\label{rel:g-om-A}
\end{equation}
The adjoint $A^{\dagger}$ is defined by $g(AX, Y) = g(X, A^{\dagger} Y)$ which, when combined with  
\eqref{rel:g-om-A}, yields $A^{\dagger} = - A$. The almost complex structure is then defined by
\begin{equation}
J_M=\left(\sqrt{AA^\dagger}\right)^{-1}A \;.
\end{equation}
In general, however, this almost complex structure is not compatible with $g_M$.
The compatible metric is
\begin{equation}
\tilde g_M(X,Y)=g_M\left(\sqrt{AA^{\dagger}}X,Y\right)= - \omega_M(X,J_MY) \;,
\end{equation}
which does not coincide with $g_M$. 

Let us perform the construction of $J_M$ for the case at hand. This can be easily done by employing
matrix notation. Let $\mathrm{F}=(F_{i\bar \jmath})$ and denote by $\mathrm{F}_+$ and $\mathrm{F}_-$
its symmetric and antisymmetric parts respectively. In terms of 
these matrices we have
\begin{equation}
g_M= \left( \begin{array}{cc}
\frac{1}{2} N & -i \mathrm{F}_+  \\
i \bar{\mathrm{F}}^T_+ & \frac{1}{2} N \end{array} \right)
\end{equation}
and
\begin{equation}
\omega_M= \left( \begin{array}{cc}
\frac{i}{2} N & \mathrm{F}_-  \\
-\bar{\mathrm{F}}^T_- & -\frac{i}{2} N \end{array} \right) \;.
\end{equation}
Now we compute $A=\omega_Mg_M^{-1}$ and obtain, by using the Schur complement,
\begin{equation}
A= \left( \begin{array}{cc}
iS_B\bar S_A^{-1} & -\mathrm{F}^TS_A^{-1}  \\
-\bar{\mathrm{F}}^T\bar S_A^{-1}&- i\bar S_BS_A^{-1} \end{array} \right),
\end{equation}
where
\begin{eqnarray}
S_A&=&\frac{N}{2} -2 \bar{\mathrm{F}}^T_+ N^{-1}\mathrm{F}_+ \xrightarrow{\rm Wil}\frac{N}{2} , \\
S_B&=&\frac{N}{2}-2\mathrm{F}_-N^{-1} \bar{\mathrm{F}}^T_+  \xrightarrow{\rm Wil}\frac{N}{2}.
\end{eqnarray}
Observe that in the Wilsonian limit ($F_{i\bar \jmath} =0$) this reduces to
\begin{equation}
A=i \left( \begin{array}{cc}
\mathbb{I} & 0  \\
0 &- \mathbb{I} \end{array}\right),
\end{equation}
and hence $AA^{\dagger} = I$,
so that 
$A=J_M$, and we get a compatible triple $(g_M,\omega_M,J_M)$, where $J_M$ is the standard almost complex structure. 

On the other hand, to first order in $\mathrm{F}$, we obtain
\begin{equation}
A= \left( \begin{array}{cc}
i \mathbb{I} & - 2 \mathrm{F}^T N^{-1} \\
- 2 \bar{\mathrm{F}}^T N^{-1}  &- i  \mathbb{I} \end{array}\right)
\label{A=J-first}
\end{equation}
as well as $AA^{\dagger} = I$.  Therefore $J_M = A$ and the compatible metric $\tilde{g}_M$ equals $g_M$.  Thus, to first
order in $\mathrm{F}$, $(g_M,\omega_M,J_M)$
forms a compatible triple.

\section{Intrinsic torsion class analysis \label{sec:int-tors}}

In this section we consider the target-space metric of the deformed sigma-model \eqref{eq:lag-nonhol-final}
and study the intrinsic torsion of the target-space following \cite{Chiossi:2002tw}.  To this end we take
$\Upsilon$ and $\bar \Upsilon$ to be constant parameters. Then, $\Psi = \Upsilon/(Y^0)^w$  
(and similarly $\bar \Psi$)
become functions of $(z, \bar z, R, \Upsilon, \bar \Upsilon)$, given by
\eqref{psi-sig-R}.  Note that in \eqref{psi-sig-R},
$\Sigma$ depends on $\Psi$ and $\bar \Psi$, so that to obtain $\Psi(z, \bar z, R, \Upsilon, \bar \Upsilon)$
one has to proceed by iteration in $\Upsilon$ and
$\bar \Upsilon$.  For instance, to first-order in $\Upsilon$, we obtain
\begin{equation}
\Psi^{(1)}(z, \bar z, R, \Upsilon) = \Upsilon \left( i \Sigma^{(0)} \, R \right)^{w/2} \;,
\label{ups-1}
\end{equation}
where $\Sigma^{(0)} = \Sigma (\Omega =0)$.  Proceeding in this way, 
the target-space metric is parametrized in terms of coordinates
$(z^i, \bar{z}^i, R)$ (with $i=1, \dots, n$). Generically, this metric does not describe the metric of a circle fibration. 
We restrict ourselves to 
a co-dimension one hypersurface $R = {\rm constant}$ and analyze the resulting geometry in terms of torsion classes.
In order to use the results of \cite{Chiossi:2002tw}, we focus on the case with $n=3$ in the following, 
and consider a six-dimensional manifold $M$ with Riemannian metric (we drop the symbol $\otimes_{\rm sym}$ in
the following)
\begin{equation}
 ds^2=g_{i\bar \jmath}dz^id \bar z^{j}+g_{ij}dz^id z^j+ g_{\bar \imath \bar \jmath }d{\bar z}^{i} 
 d {\bar z}^{j} \;\;\;,\;\;\; i,j=1,2,3 \;.
 \label{metric-6}
\end{equation}
This metric, which follows from \eqref{eq:lag-nonhol-final} by setting $\Upsilon$ and $R$ to a constant, depends on $\Omega$ 
in a rather complicated way.

We introduce complex vielbein 1-forms $e_i$,
\begin{equation}
 ds^2= \delta^{ij} \, \bar e_{\bar \imath} \, e_j.
\end{equation}
Expressing $e_i$ in terms of $dz^i$ and $d\bar z^i$ and vice-versa, 
\begin{eqnarray}
e_i &=& A_{ij}dz^j+B_{i\bar \jmath}d\bar z^{j} \nonumber\\
dz^i &=& M^{ij}e_j+N^{i\bar \jmath}\bar e_{\bar \jmath},
\label{rel-e-dz}
\end{eqnarray}
we infer the relation
\begin{equation}
\left( \begin{array}{ccc}
M & N  \\
\bar N & \bar M  \end{array} \right) = 
\left( \begin{array}{ccc}
A & B  \\
\bar B & \bar A  \end{array} \right)^{-1} \;,
\end{equation}
so that
\begin{eqnarray}
 M &=& \left(A-B\bar A^{-1}\bar B\right)^{-1} \;, \nonumber\\
  N&=& -A^{-1}B\bar M \;.
 \label{NBrel}
\end{eqnarray}
The metric coefficients can be expressed in terms of $A_{ij}$ and $B_{ij}$ as
\begin{equation} \label{full}
  g_{i\bar \jmath}=\delta_{k\bar k}\left(\bar A_{\bar k\bar \jmath}A_{ki}+\bar{ B}_{\bar ki} B_{k\bar \jmath}\right)
\end{equation}
and
\begin{equation}\label{ort} 
  g_{ij}=\frac{\delta_{k\bar k}}{2}\left(\bar {B}_{\bar ki}A_{kj}+\bar {B}_{\bar kj}A_{ki}\right).
\end{equation}
Using \eqref{ort} we express $B_{i\bar \jmath}$ in terms of $g_{ij}$ and $A_{ij}$ as
\begin{equation}
 B_{i\bar \jmath}=\delta_{i\bar p} \bar A^{\bar l\bar p} \, 
 g_{\bar l\bar \jmath} \;.
 \label{Bibj}
\end{equation}
where $\bar A^{\bar l\bar p} \bar{A}_{\bar p \bar k} = \delta^{\bar l}_{\bar p}$.

We introduce an almost complex structure, whose associated 2-form is (we use the same symbol $J$ for both 
quantities)
\begin{equation}
J= - \frac{i}{2} \, \delta^{ij} \, \bar e_{\bar \imath}\wedge e_j
\end{equation}
which, when expressed in terms of the matrices $A$ and $B$, takes the form
\begin{equation}\label{ACS2}
 J= - \frac{i}{2} \left(\bar B_{\bar \imath k}B_{i\bar l}-A_{ik}\bar A_{\bar \imath \bar l}\right) \, 
 dz^k\wedge d\bar z^{l}+i\bar B_{\bar \imath k}A_{il} \, d z^k\wedge d z^l+
 i\bar A_{\bar \imath \bar k}B_{i\bar l} \, d\bar z^{k}\wedge d\bar z^{l}.
\end{equation}
The triplet $(M, g, J)$ defines a $U(3)$ structure on $M$. An $SU(3)$ structure is obtained
by introducing a non-degenerate complex $(3,0)$-form $\psi$,
\begin{equation}
 \psi=e_1\wedge e_2\wedge e_3 \;.
\end{equation}
Equivalently, we can define an $SU(3)$ structure on $M$ as the triplet $(M, J, \psi)$, where $J$ 
is a real 2-form and $\psi$ 
 is a complex 3-form $\psi = \psi_+ + i \psi_-$ such that the relations
$J \wedge \psi_{\pm} =0 \,,\, \psi_+ \wedge \psi_- = \tfrac23 J \wedge J \wedge J \neq 0$ hold.

The failure of the holonomy group of the Levi-Civita connection of $g$ to reduce to 
$SU(3)$ is measured by the so-called intrinsic torsion $\tau$. 
The space to which $\tau$ belongs can be decomposed into five classes \cite{Gray:1980,Chiossi:2002tw},
\begin{equation}
 \tau\in{\cal W}_1\oplus{\cal W}_2\oplus{\cal W}_3\oplus{\cal W}_4\oplus{\cal W}_5 \;.
\end{equation}
The five torsion classes of $(M, J, \psi)$ are defined through the decomposition of the exterior
derivative of $J, \psi$ into $SU(3)$-modules, and they describe the failure
of $J, \psi$ to being closed, as follows \cite{Chiossi:2002tw},
\begin{eqnarray}
&{\cal W}_1& \leftrightarrow \left(dJ\right)^{(3,0)}\;,
\;\;\;\;\;\;\;\; {\cal W}_2 \leftrightarrow \left(d\psi\right)_0^{(2,2)} \;, \nonumber\\
&{\cal W}_3& \leftrightarrow \left(dJ\right)_0^{(2,1) }\;,
\;\;\;\;\;\;\;\; {\cal W}_4 \leftrightarrow J\wedge dJ \;, \nonumber\\
&{\cal W}_5& \leftrightarrow \left(d\psi\right)^{(3,1)} \;.
\end{eqnarray}
The subscript $0$ means that only the primitive part of the form should be retained, 
i.e. forms that are in  the kernel of $(J\wedge\bullet)$. 
Thus, $\beta \in \Lambda_0^{(2,2)}$ if $J \wedge \beta =0$, and $\gamma \in
\Lambda_0^{(2,1)}$ if $J \wedge \gamma =0$.  The vanishing of both ${\cal W}_1$ and ${\cal W}_2$
implies that the manifold is hermitian, while the vanishing of the four torsion classes ${\cal W}_1, \dots, {\cal W}_4$ implies that the manifold is K\"ahler.

We proceed to determine the torsion classes.  To do so, we write $d e_i$ as 
\begin{equation}
de_i=\left({\cal O}_1\right)_i^{\; mn}e_m\wedge e_n+\left({\cal O}_2\right)_i^{\; \bar mn} \bar e_{\bar m}\wedge e_n+
\left({\cal O}_3\right)_i^{\;\bar m\bar n}\bar e_{\bar m}\wedge\bar e_{\bar n}.
\label{O-coeff}
\end{equation}
The torsion classes are then expressed as
\begin{eqnarray}
&{\cal W}_1& \leftrightarrow \bar{\cal Z}^{jk\bar \jmath \bar k} \,
\psi \;,
\nonumber\\
&\tilde {\cal W}_2& \leftrightarrow {\cal Z}^{\bar m\bar njk}\bar e_{\bar m\bar n}\wedge e_{jk} \;,
\nonumber\\
&\tilde {\cal W}_3& \leftrightarrow {\cal C}^{ij\bar k} e_{ij}\wedge \bar e_{\bar k} \;,
\nonumber\\
&{\cal W}_4& \leftrightarrow {\cal R}^{\bar n\bar k} \, \psi\wedge \bar e_{\bar n \bar k}+{\cal S}^{nk}
\, \bar\psi\wedge  e_{nk} \;,
\nonumber\\
&{\cal W}_5& \leftrightarrow \left({\cal O}_2\right)_i^{\; \bar mi} \bar e_{\bar m}\wedge \psi \;,
\label{torsion-O}
\end{eqnarray}
where $e_{jk} = e_j \wedge e_k$, $\epsilon_{ijk}$ denotes the totally anti-symmetric epsilon tensor, and where
we introduced the following tensors,
\begin{eqnarray}
 {\cal C}^{ij\bar k}&=&-\frac{1}{2}\left[ \left(\delta^{\bar l j}\left(\bar {\cal O}_2\right)_{\bar l}^{\; i\bar k}+\delta^{\bar k l}\left({\cal O}_1\right)_l^{\; ij}  \right)-\left(i\leftrightarrow j\right) \right]\;,
 \nonumber\\
 {\cal E}^{i\bar \jmath \bar k}= - \bar{\cal C}^{\bar \jmath \bar k i} &=& 
 \frac{1}{2} \left[
 \left(\delta^{\bar l i}\left(\bar {\cal O}_1\right)_{\bar l}^{\; \bar \jmath\bar k}-\delta^{\bar \jmath l}\left({\cal O}_2\right)_l^{\;\bar ki} \right)-\left(\bar \jmath\leftrightarrow \bar k\right) \right] \;,\nonumber\\
 {\cal R}^{\bar n\bar k}&=&-\frac{1}{2}\left[ \delta^{\bar n l}{\cal C}^{ij\bar k}\epsilon_{lij}-\left(\bar n\leftrightarrow \bar k\right) \right] \;,\nonumber\\
 {\cal S}^{nk} = \overline{\left({\cal R}^{\bar n\bar k} \right)}
 &=&\frac{1}{2}\left[ \delta^{n \bar m}{\cal E}^{k\bar \jmath \bar l}\epsilon_{\bar m\bar \jmath \bar l}-\left(n\leftrightarrow k\right) \right] \;,\nonumber\\
 {\cal Z}^{\bar m\bar njk}&=&\left({\cal O}_3\right)_i^{\;\bar m\bar n}\epsilon^{ijk}.
 \label{W-tensors}
\end{eqnarray}
Note that in \eqref{torsion-O} we used $\tilde {\cal W}$ on the left hand side of some of the equations, to 
indicate that in order to obtain 
the primitive part of the forms, we have to impose the following conditions.
The torsion class ${\cal W}_2$ has to satisfy $J \wedge {\cal W}_2 = 0$. We compute
\begin{equation}
 J \wedge \tilde{\cal W}_2 \propto \left( {\cal Z}^{\bar \jmath \bar kjk}-{\cal Z}^{\bar k\bar \jmath jk}
 \right) {\bar \psi} \wedge  \psi = 2 {\cal Z}^{\bar \jmath \bar kjk} \, {\bar \psi} \wedge  \psi  \;,
\end{equation}
which is proportional to the torsion class ${\cal W}_1$.
Thus, we conclude that when ${\cal W}_1$ vanishes, $\left(d\psi\right)^{(2,2)}$ is primitive, i.e.
$\left(d\psi\right)^{(2,2)} = \left(d\psi\right)_0^{(2,2)}$, and $\tilde{\cal W}_2= {\cal W}_2$.

The torsion  class ${\cal W}_3$, on the other hand, has to satisfy $J \wedge {\cal W}_3 = 0$.  We compute  
\begin{equation}
J \wedge \tilde{\cal W}_3 = {\cal C}^{ij\bar k} \epsilon_ {ij l } \,  \delta^{lp} \, \psi 
\wedge {\bar e}_{\bar k \bar p} \;,
\end{equation}
which is proportional to the torsion class ${\cal W}_4$.  Thus, when ${\cal W}_4$ vanishes, we have
$\tilde{\cal W}_3 = {\cal W}_3$.

The coefficients $\left({\cal O}\right)$ appearing in \eqref{W-tensors}
can be expressed
in terms of the matrices $A,B, M$ and $N$ introduced in \eqref{rel-e-dz}.  We obtain
\begin{eqnarray}
\label{O3}
 \left({\cal O}_1\right)_i^{\;mn}&=&\frac{1}{2}\left(\left(\partial_k A_{ij}M^{km}+\partial_{\bar k}A_{ij}\bar N^{\bar km}\right)M^{jn}-\left(m\leftrightarrow n\right)\right) \nonumber\\
 &&+\frac{1}{2}\left(\left(\partial_k B_{i\bar \jmath}M^{km}+\partial_{\bar k}B_{i\bar \jmath}\bar N^{\bar km}\right)\bar N^{\bar \jmath n}-\left(m\leftrightarrow n\right)\right) \;, \nonumber\\
  \left({\cal O}_2\right)_i^{\;\bar mn}&=&\partial_kA_{ij}\left(M^{jn}N^{k\bar m}-M^{kn}N^{j\bar m}\right)+\partial_{\bar k}A_{ij}\left(\bar M^{\bar k\bar m}M^{jn}-\bar N^{\bar kn}N^{j\bar m}\right) \nonumber\\
&&+ \partial_k B_{i\bar \jmath}\left(N^{k\bar m}\bar N^{\bar \jmath n}-M^{kn} \bar M^{\bar \jmath \bar m}\right)+\partial_{\bar k}B_{i\bar \jmath}\left(\bar M^{km}\bar N^{\bar \jmath n}-\bar N^{\bar kn}\bar M^{\bar \jmath \bar m}\right) \;, \nonumber\\
 \left({\cal O}_3\right)_i^{\;\bar m\bar n}&=&\frac{1}{2}\left(\left(\partial_kA_{ij}N^{k\bar m}+\partial_{\bar k}A_{ij}\bar M^{\bar k\bar m}\right)N^{j\bar n}
   -\left(\bar m\leftrightarrow\bar n\right)\right) \nonumber\\
   &&+\frac{1}{2}\left(\left( \partial_kB_{i\bar \jmath}N^{k\bar m}+\partial_{\bar k}B_{i\bar \jmath}\bar M^{\bar k\bar m}\right)\bar M^{\bar \jmath \bar n} -\left(\bar m\leftrightarrow\bar n\right)\right).
\end{eqnarray}

There are two cases to consider, namely $B_{i \bar \jmath}=0$ and $B_{i \bar \jmath}\neq 0 $.
They can be understood as follows.
The metric \eqref{metric-6}, which is obtained from \eqref{eq:lag-nonhol-final},
depends on $\Omega$ in a rather complicated way.  We may expand
the metric coefficients in powers of $\Omega$ and compute the torsion classes order
by order in $\Omega$.  When $\Omega =0$,
we have $g_{ij} = 0$ and we infer from \eqref{Bibj} that $B_{i \bar \jmath } =0$.
In this case the metric is K\"ahler, 
since it corresponds to the sigma-model metric in the absence of a chiral background. On the other hand,
when $\Omega \neq 0$ we have $g_{ij} \neq 0$, as discussed at the end of subsection \ref{sec:wils},
and hence also $B_{i \bar \jmath } \neq 0$.

We first consider the case when $B_{i \bar \jmath } =0$ and verify that the vanishing of the first four torsion classes 
implies that the manifold is K\"ahler.
We infer from \eqref{NBrel} that $N^{i \bar \jmath} =0$, and from 
\eqref{ort} and \eqref{full} we obtain
$g_{ij}=0$ and $g_{i\bar \jmath}=\delta_{k\bar k}A_{ki}\bar A_{\bar k\bar \jmath}$. It follows that
$\left({\cal O}_3\right)=0$, 
so that ${\cal W}_ 1={\cal W}_2=0$, and hence the 
manifold is hermitian. For  $\left({\cal O}_1\right)$ and $\left({\cal O}_2\right)$ we obtain
\begin{equation}
\left({\cal O}_1\right)_i^{\; mn}=\frac{1}{2}\left(\partial_kA_{ij}M^{km}M^{jn}-(m\leftrightarrow n)\right)
\end{equation}
and
\begin{equation}
\left({\cal O}_2\right)_i^{\; \bar mn}=\partial_{\bar k}A_{ij}\bar M^{\bar k\bar m}M^{jn}.
\end{equation}
This results in 
\begin{equation}
 {\cal C}^{ij\bar k}=- \frac12 \left[ \partial_n g_{l\bar m} M^{lj}M^{ni}\bar M^{\bar m\bar k} - ( i \leftrightarrow j)
 \right] \;,
\end{equation}
which vanishes iff
$\partial_{n}g_{l \bar m }=\partial_{l}g_{n \bar m}$, in which case
\begin{equation}
 {\cal W}_ 3={\cal W}_ 4=0.
\end{equation}
Thus we recover the well known fact that the vanishing of the first four torsion classes 
implies that the manifold is K\"ahler.

We now briefly discuss the torsion class 
${\cal W}_5$ when $B_{i \bar \jmath } =0$. In this case ${\cal W}_5$ is  
proportional to
\begin{equation}
\left({\cal O}_2\right)_i^{\; \bar mi}=\partial_{\bar k}A_{ij}\bar M^{\bar k\bar m}M^{ji} \;.
\end{equation}
Taking $M^{ij}$ to be diagonal, 
\begin{equation}
M^{ij} = \lambda_i(z, \bar z) \, \delta^{ij} \;\;\;,\;\;\; \lambda_i \in \mathbb{C} \;,
\end{equation}
we have $\ln \sqrt{\det g_{i \bar \jmath}} = - \frac12 \ln \prod_{i=1}^3 \left( \lambda_i {\bar \lambda}_i \right)$,
and
we obtain (no summation over $\bar m$ here)
\begin{equation}
 \left({\cal O}_2\right)_i^{\; \bar mi}=-\sum_{i=1}^{3}\frac{\bar\lambda_{\bar m}}{\lambda_i}\partial_{\bar m}\lambda_i
 = - \bar\lambda_{\bar m} \partial_{\bar m} \ln \prod_{i=1}^3 \lambda_i \;.
 \label{cond-w5}
 \end{equation}
In order for ${\cal W}_5$ to vanish we must have
\begin{equation}
\partial_{\bar m} \ln \prod_{i=1}^3 \lambda_i =0 \;\;\;,\;\;\; m =1,2,3 \;,
\end{equation}
which, together with the complex-conjugate equation, 
implies the vanishing  
the first Chern class 
\begin{equation}
 c_1=\left[ \frac{1}{2\pi}{\cal R}\right] \;,
\end{equation}
where ${\cal R}$ is the Ricci form
\begin{equation}
 {\cal R}=i\partial\bar\partial\ln \sqrt{\det g_{i \bar \jmath}} \;.
\end{equation}

Next, we consider the case when  $B_{i \bar \jmath }\neq 0$, in which case $g_{ij}\neq0$.
We may expand
the metric coefficients in powers of $\Omega$.
In the following, we compute the four torsion classes ${\cal W}_1, \dots, {\cal W}_4$ to first
order in $\Omega$.   
At first order in $\Omega$ we have $g_{ij}\neq0$ and hence, 
according to \eqref{Bibj}, this  
means that we will work at first order in $B_{i \bar \jmath }$.  
We therefore expand the metric \eqref{metric-6} in powers of $B$ as 
\begin{equation}
 ds^2=\left( g^{(0)}_{i\bar \jmath} +  g^{(1)}_{i\bar \jmath}\right)
 dz^id \bar z^{j}+ g^{(1)}_{ij} 
  dz^id z^j+  
  g^{(1)}_{\bar \imath \bar \jmath } d{\bar z}^{i} 
 d {\bar z}^{j} \;,
 \label{metric-6-exp}
\end{equation}
where the superscript indicates the power of $B$ (or, equivalently, the power of $\Omega$).  
For later use, we introduce the Christoffel symbol associated to the K\"ahler metric $g^{(0)}$,
\begin{equation}
\Gamma_{\bar k\bar l}^{\bar p}= g^{(0) {\bar p r}} \partial_{\bar k}g^{(0)}_{r\bar l}.
\label{christ}
\end{equation}

We now compute
the coefficients $\left( {\cal O} \right)$ given in \eqref{O3}
to first order in  $B_{i\bar \jmath}$. Observe that $N^{i \bar \jmath}$
is then of first order in $B$.
First, we rewrite $ \left({\cal O}_3\right)$ given in \eqref{O3} as 
\begin{equation}
  \left({\cal O}_3\right)_i^{\;\bar m\bar n}=\frac{1}{2}\left[ A_{il}\bar M^{\bar \jmath \bar n}\left(N^{k\bar m}\partial_k\left(A^{ls}B_{s\bar \jmath}\right)+
   \bar M^{\bar k\bar m}\partial_{\bar k}\left(A^{ls}B_{s\bar \jmath}\right)\right)-\left(\bar m\leftrightarrow\bar n\right)\right] \;.
\end{equation}
To first order in $B_{i \bar \jmath}$ this yields,
\begin{equation}
  \left({\cal O}_3\right)_i^{\;\bar m\bar n}=\frac{A_{il}^{(0)}}{2}\left(\bar M_0^{\bar \jmath\bar n}\bar M_0^{\bar k\bar m}\partial_{\bar k}\left(g^{(0) \bar p l} \, g_{\bar p\bar \jmath}^{(1)}\right)
-\left(\bar m\leftrightarrow\bar n\right)\right),
\end{equation}
where we used \eqref{Bibj}, and where 
$g^{(0) \bar p l } = {\bar A}^{\bar p \bar k} A^{lk}$. Using \eqref{christ}
we have
 \begin{equation}
  \left({\cal O}_3\right)_i^{\;\bar m\bar n}=\frac{\delta_{i\bar \imath}}{2}\bar M_0^{\bar \jmath\bar n}\bar M_0^{\bar k\bar m}\bar M_0^{\bar l\bar \imath}\left(\partial_{\bar k} g_{\bar l\bar \jmath}^{(1)}-
   \Gamma_{\;\bar k\bar l}^{\bar p} g_{\bar p \bar \jmath}^{(1)}-\left(\bar \jmath \leftrightarrow\bar k\right)\right).
 \end{equation}
Introducing the covariant derivative with respect to the Levi-Civita connection \eqref{christ},
\begin{equation}
 {\cal D}_{\bar k} g_{\bar l\bar \jmath}^{(1)}=\partial_{\bar k} g_{\bar l\bar \jmath}^{(1)}-
\Gamma_{\bar k\bar l}^{\bar p} g_{\bar p\bar \jmath}^{(1)}-\Gamma_{\bar k \bar \jmath}^{\bar p} g_{\bar l\bar p}^{(1)}\;,
\end{equation}
we obtain
 \begin{eqnarray}\label{O3cov}
  \left({\cal O}_3\right)_i^{\;\bar m\bar n}
  =\frac{\delta_{i\bar \imath}}{2}\bar M_0^{\bar \jmath\bar n}\bar M_0^{\bar k\bar m}\bar M_0^{\bar l\bar \imath}
  \left( {\cal D}_{\bar k} g_{\bar l\bar \jmath}^{(1)}-{\cal D}_{\bar \jmath} g_{\bar l\bar k}^{(1)}\right) \;.
  \end{eqnarray}
It follows that to first order in $\Omega$, 
\begin{eqnarray}
\bar{\cal W}_1 \propto \left({\cal O}_3\right)_i^{\;\bar m\bar n} \epsilon^{i m n} = 0 \;,
\label{van-w1}
 \end{eqnarray}
and hence $\tilde{{\cal W}}_2 ={\cal W}_2$.

Similarly, we compute 
\begin{equation}
 {\cal C}^{ij\bar k}=-\frac{1}{2}\bar M_0^{\bar m\bar k}M_0^{li}M_0^{nj}\left({\cal D}_lg_{n\bar m}^{(1)}-{\cal D}_ng_{l\bar m}^{(1)} \right) \;.
 \label{C-Om}
\end{equation}
This determines $\tilde{\cal W}_3$ and ${\cal W}_4$ to first order in $\Omega$.
Notice that \eqref{C-Om} vanishes provided that 
$\partial_l g_{n\bar m}^{(1)}=\partial_n g_{l\bar m}^{(1)}$, i.e. if $g_{n\bar m}^{(1)}$ is K\"{a}hler.

Summarizing, at first order in $\Omega$ we find that ${\cal W}_1 =0$, while ${\cal W}_2$ is given by \eqref{O3cov}.
A non-vanishing class ${\cal W}_2$ implies that the almost complex structure is not integrable.
The torsion classes $\tilde{\cal W}_3$ and ${\cal W}_4$ are determined by \eqref{C-Om}.

In the next section, we give an example of a deformed sigma-model metric that, to first order in $\Omega$,
has ${\cal W}_1 = {\cal W}_3 = {\cal W}_4 =0$. Thus, the target manifold is almost-K\"ahler, since $dJ=0$ 
\cite{Gray:1980}.

\section{Example \label{sec:examp}}

In the following we identify the chiral background with the square of the Weyl multiplet,
so that the model we will examine in this section describes the coupling to $R^2$-terms.
Thus the Weyl weight $w$ equals $w=2$.

The model we consider is 
based on a function $F(Y, \bar Y, \Upsilon, \bar \Upsilon)$ that is homogeneous of degree two and 
specified by (see \eqref{eq:F-0-Om})
\begin{eqnarray}
F^{(0)}(Y) &=& - \left( Y^0 \right)^2 \, z^1 z^2 z^3 \;, \nonumber\\
\Omega (Y, \bar Y, \Upsilon, \bar \Upsilon) &=& \Upsilon \, g(z) + \bar \Upsilon \, \bar g ( \bar z ) + \left( \Upsilon + \bar \Upsilon \right) \, h(z, \bar z) \;,
\label{F-STU-tri}
\end{eqnarray}
where $h$ is a real function, while $g$ is holomorphic, and 
\begin{eqnarray}
S = - i z^1 = -i \, \frac{Y^1}{Y^0} \;\;\;,\;\;\; T = - i z^2 = -i \, \frac{Y^2}{Y^0} \;\;\;,\;\;\;
U = - i z^3 = -i \, \frac{Y^3}{Y^0} \;.
\end{eqnarray}
The harmonic part of $\Omega$ determines
the Wilsonian Lagrangian.
We take $g(z)$ and $h(z, \bar z)$ to be decomposable as
\begin{eqnarray}
g(z) = \sum_{i=1}^3 f^{(i)} (z^i) \;\;\;,\;\;\;
h(z, \bar z) = \sum_{i=1}^3 p^{(i)}(z^i, \bar z^i) \;,
\label{f-p-dec}
\end{eqnarray}
with real $p^{(i)}$. Then, 
the model describes an $N=2$ STU-type model 
\cite{Sen:1995ff}
in the presence of $R^2$-interactions \cite{Gregori:1999ns}.  $\Omega$ may also receive
corrections that are of higher order in $\Upsilon$ and $\bar \Upsilon$.  Here, we will work to first order
in $\Upsilon$ and $\bar \Upsilon$.  We may therefore identify first-order in $\Upsilon$ with first-order in $\Omega$,
which is the order employed in the torsion analysis of the previous section.

Inserting \eqref{F-STU-tri} into \eqref{eq:lag-nonhol-final} results in a rather complicated target-space metric which, to first order
in $\Upsilon$, contains $d \Psi, d\bar \Psi$ and $d R$-legs.  To simplify the metric, we perform a coordinate
transformation whose form can be motivated by duality symmetry considerations, as follows.

Suppose we demand the model to have S- and T-duality symmetries.  In the presence of a chiral background,
the S- and T-duality transformation rules get modified and 
involve derivatives of $\Omega$.  For instance, under
S-duality, the fields $Y^0, S, T, U$ transform as follows \cite{Cardoso:2008fr},
\begin{eqnarray}
&& S \rightarrow \frac{aS - i b}{ic S + d} \;\;\;,\;\;\; Y^0 \rightarrow \Delta_{\rm S} \, Y^0 \;\;\;,\;\;\;
\Delta_{\rm S} = ic S + d \;,\nonumber\\
&& T \rightarrow T + \frac{2ic}{\Delta_{\rm S} (Y^0)^2} \frac{\partial
\Omega}{\partial U} \;\;\;,\;\;\;
U \rightarrow U + \frac{2ic}{\Delta_{\rm S} (Y^0)^2} \frac{\partial
\Omega}{\partial T} \;,
\label{S-T-U-transf-beh}
\end{eqnarray}
where $a, b, c, d$ are integer-valued parameters that satisfy $ad - bc =1$ 
and parametrize (a subgroup of) ${\rm SL} (2, \mathbb{Z})$.  Here we recall that $\Upsilon$ does not
transform, since it is invariant under symplectic transformations.
We obtain similar transformation rules under T- and U-duality if we assume that the model has triality symmetry.
These transformation rules then follow from \eqref{S-T-U-transf-beh} by appropriately interchanging $S, T$ and $U$.
Furthermore, for the model to possess S- and T-dualities, we need to take $\Upsilon$ to be real.  Then, at first
order in $\Upsilon$, the model based on \eqref{F-STU-tri}
is invariant under S-duality transformations 
\eqref{S-T-U-transf-beh} provided that the derivatives of $\Omega$ with respect to $S, T, U$
transform as 
\cite{Cardoso:2008fr}
\begin{eqnarray}
\left(\frac{\partial \Omega}{\partial S}\right)' = \Delta_{\rm S}^2 \, \frac{\partial \Omega}{\partial S} 
\;\;\;,\;\;\;
\left(\frac{\partial \Omega}{\partial T}\right)' = \frac{\partial \Omega}{\partial T} \;\;\;,\;\;\;
\left(\frac{\partial \Omega}{\partial U}\right)' = \frac{\partial \Omega}{\partial U} \;.
\label{dif-Om}
\end{eqnarray}
At higher order in $\Upsilon$, these expressions get corrected according to 
\cite{Cardoso:2008fr}.  Similar considerations hold for T-duality transformations.
In general, S- and T-duality invariance can only be achieved if
the real functions $p^{(i)}$ in \eqref{f-p-dec} are chosen in a suitable way.

Now we redefine the field $S$ in such a way that the new field $\tilde{S}$ transforms 
in the usual way under  S-duality \cite{Cardoso:2010gc},
\begin{equation}
\tilde{S} \rightarrow \frac{a \tilde{S} - i b}{ic \tilde{S} + d} \;,
\label{transf-tilde-S}
\end{equation}
and is invariant under T-dualities.  To first order in $\Upsilon$, this is achieved by
the following combination,\footnote{We note that the combination introduced in \cite{Cardoso:2010gc}
differs from \eqref{tilde-S} by a term that is invariant under T-dualities,
\begin{equation}
- 2 \,\frac{(S + \bar S)}{|Y^0|^2} \, \frac{\partial_{\bar S} \Omega }{(T + \bar T) ( U + \bar U)} \;.
\end{equation}
Under S-duality, this term transforms in the same way as the difference 
$\tilde{S} - S$ in \eqref{tilde-S}.}
\begin{equation}
\tilde{S} = S + \frac{2}{(Y^0)^2} \left( \frac{\partial_T \Omega}{U + \bar U} +
\frac{\partial_U \Omega}{T + \bar T} \right) \;.
\label{tilde-S}
\end{equation}
Using \eqref{S-T-U-transf-beh} and \eqref{dif-Om}
it can be readily checked that \eqref{tilde-S} transforms according to \eqref{transf-tilde-S}.
Similarly, it can be verified that $\tilde{S}$ 
is invariant under T- and U-duality transformations. At first order in $\Upsilon$, the
transformation laws for T- and U-duality are obtained from \eqref{S-T-U-transf-beh}
by interchanging $S$ with $T$ and $U$, respectively.

We also introduce new fields $\tilde{T}$ and $\tilde{U}$ that are defined as in 
\eqref{tilde-S} by interchanging $S$ with $T$ and $U$, respectively.

Having motivated the coordinate transformation \eqref{tilde-S} by using duality symmetry
considerations, we apply it to the model \eqref{F-STU-tri} and obtain
\begin{equation}
\tilde{z}^1 = z^1 - 2 i \Psi \left[ \frac{g_2 + h_2}{z^3 - \bar{z}^3} +  \frac{g_3 + h_3}{z^2 - \bar{z}^2}
\right] - 2i R^2 \, {\bar \Psi} \left[ \frac{h_2}{z^3 - \bar{z}^3} +  \frac{h_3}{z^2 - \bar{z}^2}
\right] \;,
\end{equation}
 where
 $g_2 = \partial g/\partial z^2, h_2 = \partial h/\partial z^2$, etc.  Similar expressions hold for
 $\tilde{z}^2$ and $\tilde{z}^3$.  Then, expressing the target-space metric  
 \eqref{eq:lag-nonhol-final} in terms
 of the new coordinates $\tilde{z}^i$, we obtain that to first order in $\Upsilon$, the 
 target-space metric takes the form \eqref{metric-6-exp}, with no extra $d\Psi, d \bar \Psi$ and $dR$-legs.  The metric \eqref{metric-6-exp} is now expressed in terms of the new coordinates.
 Its metric coefficients depend on $\tilde{z}^i, \bar{\tilde{z}}^i, \Psi, \bar \Psi$ and $R$.  Using
 \eqref{f-p-dec}  we find
  \begin{eqnarray}
 g^{(0)}_{i \bar \jmath} = 
 \frac{\delta_{i \bar \jmath}}{2 (\tilde{z}^i - \bar{\tilde{z}}^i)^2} \;\;\;,\;\;\; g^{(1)}_{i \bar \jmath} =0 \;,
 \end{eqnarray}
 while the metric components $g^{(1)}_{ij}$ are off-diagonal and take the form 
 \begin{equation}
 g^{(1)}_{12} = - \frac{2 i}{(\bar Y^0)^2}  \, \frac{\partial_{\bar 3} \Omega}{(\tilde{z}^1 - 
 \bar{\tilde{z}}^1)^2 (\tilde{z}^2 - 
 \bar{\tilde{z}}^2)^2 } = - 2 i\, \frac{ \left[ \bar \Psi ( \bar g_{\bar 3} + h_{\bar 3} ) + {\bar R}^2 \Psi h_{\bar 3}
 \right]
  }{(\tilde{z}^1 - 
 \bar{\tilde{z}}^1)^2 (\tilde{z}^2 - 
 \bar{\tilde{z}}^2)^2 }
 \;,
 \label{g1-off}
 \end{equation}
and similarly for the components $g^{(1)}_{13}$ and  $g^{(1)}_{23}$, which follow from \eqref{g1-off} by suitably
interchanging
$\tilde{z}^1, \tilde{z}^2$ and $\tilde{z}^3$.  
Observe that when the model has S- and T-duality symmetries, 
the combination \eqref{g1-off} is such that $g^{(1)}_{ij} d \tilde{z}^i d \tilde{z}^j$ is
duality invariant at first order in $\Upsilon$.

 Now we return to the torsion analysis.   Using the above target-space metric, we immediately
 infer that \eqref{C-Om} vanishes, so that ${\cal W}_3 = {\cal W}_4 =0$.  On the other hand, using 
\eqref{torsion-O}, \eqref{W-tensors} and 
\eqref{O3cov} we find that the torsion 4-form $\tilde{\cal W}_2$ is proportional to 
 \begin{eqnarray}
 \tilde{\cal W}_2 \propto \,\bar{e}_{\bar m \bar n} \wedge 
 \left(e_{23} 
 \, {\cal D}_{\bar m} g^{(1)}_{\bar 1 \bar n} + e_{31} 
 \, {\cal D}_{\bar m} g^{(1)}_{\bar 2 \bar n} + e_{12} \, {\cal D}_{\bar m} g^{(1)}_{\bar 3 \bar n} 
 \right) \;.
 \label{w2-til-exp}
 \end{eqnarray}
Focussing on the terms proportional to $e_{23}$ we obtain
\begin{eqnarray}
e_{23} \wedge \left({\bar e}_{\bar 1 \bar 2} \, {\cal D}_{\bar 1} g^{(1)}_{\bar 1 \bar 2}
+ \bar{e}_{\bar 1 \bar 3} \, {\cal D}_{\bar 1} g^{(1)}_{\bar 1 \bar 3} + 
\bar{e}_{\bar 2 \bar 3} \left( {\cal D}_{\bar 2} g^{(1)}_{\bar 1 \bar 3} - 
{\cal D}_{\bar 3} g^{(1)}_{\bar 1 \bar 2} \right) \right) \;,
\label{e23-ter}
\end{eqnarray}
where we used that the metric coefficients $g^{(1)}_{\bar \imath \bar \jmath}$ are off-diagonal.  Similar expressions
hold for the other terms in \eqref{w2-til-exp}.  Using these, we verify that $\tilde{\cal W}_2 \wedge J =0$,
which implies that $\tilde{\cal W}_2 = {\cal W}_2$, as was expected since ${\cal W}_1=0$ (see \eqref{van-w1}).
Evaluating \eqref{e23-ter} using \eqref{f-p-dec} we obtain
\begin{eqnarray}
&& 2 i \, e_{23} \wedge \left[ \frac{
 \left( \Psi + R^2 \bar \Psi \right) 
{\bar e}_{\bar 3 \bar 2} \, h_{\bar 3 3} 
}{(\tilde{z}^1 - 
 \bar{\tilde{z}}^1)^2 (\tilde{z}^2 - 
 \bar{\tilde{z}}^2)^2 } +  
 \frac{
 \left( \Psi + R^2 \bar \Psi \right) 
{\bar e}_{\bar 2 \bar 3} \, h_{\bar 2 2} 
}{(\tilde{z}^1 - 
 \bar{\tilde{z}}^1)^2 (\tilde{z}^3 - 
 \bar{\tilde{z}}^3)^2 } \right] \\
&& +  2 i \, e_{23} \wedge \frac{
{\bar e}_{\bar 1 \bar 2}\, \left[
\partial_{\bar 1} \left(
\Psi + R^2 \bar \Psi \right) h_3 + \left(\partial_{\bar 1} \Psi\right) g_3 \right]
+ {\bar e}_{\bar 3 \bar 2}\, \left[
\partial_{\bar 3} \left(
\Psi + R^2 \bar \Psi \right) h_3 + \left(\partial_{\bar 3} \Psi\right) g_3 \right]
}{(\tilde{z}^1 - 
 \bar{\tilde{z}}^1)^2 (\tilde{z}^2 - 
 \bar{\tilde{z}}^2)^2 } \nonumber\\
&& + 
 2 i \, e_{23} \wedge \frac{
 {\bar e}_{\bar 1 \bar 3}\, \left[
\partial_{\bar 1} \left(
\Psi + R^2 \bar \Psi \right) h_2 + \left(\partial_{\bar 1} \Psi\right) g_2 \right]
+ {\bar e}_{\bar 2 \bar 3}\, \left[
\partial_{\bar 2} \left(
\Psi + R^2 \bar \Psi \right) h_2 + \left(\partial_{\bar 2} \Psi\right) g_2 \right]
}{(\tilde{z}^1 - 
 \bar{\tilde{z}}^1)^2 (\tilde{z}^3 - 
 \bar{\tilde{z}}^3)^2 } \;. \nonumber
 \label{comb-w2}
 \end{eqnarray}
This vanishes in the Wilsonian limit provided we take $\Psi$ to be an independent field, in agreement with the discussion
presented in subsection \ref{sec:wils}. On the other hand, when turning on non-holomorphic terms (which are encoded in $h(z, \bar z)$), the combination \eqref{comb-w2} is non-vanishing.  Similar considerations hold for the other terms in
\eqref{w2-til-exp}.  
Thus we conclude that the torsion class
${\cal W}_2$ becomes non-vanishing in the presence of non-holomorphic terms and/or when taking $\Psi$ 
to be a dependent field, as in \eqref{ups-1}.
Finally, we note that also the torsion class ${\cal W}_5$ is non-vanishing for this model,
since \eqref{cond-w5} is already non-vanishing when $\Omega=0$.

\section{Real coordinates and the Hesse potential \label{sec:hessep}}

In \cite{Cardoso:2012nh} a theorem was presented according to which an arbitrary point-particle Lagrangian 
${\cal L}(\phi, \dot \phi)$
depending on coordinates $\phi^i$ and velocities $\dot \phi^i$
can be formulated in terms of a complex function $F(x, \bar x)$, where $x^i = \tfrac12 (\phi^i + i \dot \phi^i)$,
such that the canonical variables $(\phi^i, \pi_i = \partial {\cal L}/\partial \dot \phi^i)$
equal 
\begin{eqnarray}
\phi^i &=& 2 \, {\rm Re} \, x^i \;, \nonumber\\
\pi_i & = & 2 \, {\rm Re} \, F_i \;\;\;,\;\;\; F_i =  \frac{\partial F(x, \bar x)}{\partial x^i} \;.
\label{phi-pi}
\end{eqnarray}
The function $F(x, \bar x)$ takes the form displayed in \eqref{eq:F-0-Om}, and the Lagrangian ${\cal L}$ and the Hamiltonian $H$
are expressed in terms of $F$ as
\begin{eqnarray}
\label{L-H}
{\cal L} &=& 4 \left( {\rm Im }\, F - \Omega \right) \;,\\
H &=& {\dot \phi}^i \, \pi_i - {\cal L}(\phi, \dot \phi) = 
- i \left( x^i \, {\bar F}_{\bar \imath} - \bar{x}^i \, F_i \right) - 4 \, {\rm Im} \left( F^{(0)} - 
\frac12 x^i \, F_i^{(0)} \right) - 2 \left( 2 \Omega - x^i \, \Omega_i - \bar{x}^i \, \Omega_{\bar \imath} \right) \;,
\nonumber
\end{eqnarray}
where $F^{(0)}_i = \partial F^{(0)}/\partial x^i, \Omega_i = \partial \Omega/\partial x^i$, and similarly
for their complex conjugates.

In the absence of a chiral background, it is known that the Hesse potential underlying
the 
real formulation of special geometry \cite{Freed:1997dp,Cortes:2001qd,Ferrara:2006js,Ferrara:2006at,Mohaupt:2011aa} 
equals the 
Hamiltonian $H$ displayed above (which in this case reads $H = {\dot \phi}^i \, \pi_i -  4 \, {\rm Im } \,F^{(0)}$).
This continuous to hold in the 
presence of a chiral background, since $H$ given in \eqref{L-H} incorporates the chiral background in a manner
that respects the symplectic nature of $H$.  We proceed to verify this explicitly. We follow the exposition of 
\cite{Mohaupt:2011aa}.

We consider the function $F(Y, \bar Y, \Upsilon, \bar \Upsilon)$ given in \eqref{eq:F-0-Om}, which is 
homogeneous of degree two.
We follow the theorem outlined above and 
introduce real coordinates\footnote{The $x^I$ introduced here should not be confused with the 
$x^i$ in \eqref{phi-pi}, which are complex.} $(x^I, y_I)$
\begin{eqnarray}
Y^I = x^I + i u^I (x,y) \;\;\;,\;\;\; F_I= y_I + i v_I (x,y) \;.
\label{eq:darboux}
\end{eqnarray}
as well as
\begin{eqnarray}
{\cal L}(Y, \bar Y, \Upsilon, \bar \Upsilon) &=& {\rm Im} \, F - \Omega \;, \nonumber\\
H(x, y, \Upsilon, \bar \Upsilon) &=& u^I \, y_I - {\cal L} \;.
\label{eq:L-aux}
\end{eqnarray}
Here, we have scaled out a factor $4$ relative to \eqref{L-H}.

We introduce the symplectic vector 
\begin{equation}
 V_a=\begin{pmatrix} Y^I \\F_I(Y, \bar Y, \Upsilon, \bar \Upsilon) \end{pmatrix}, 
\end{equation}
and decompose it into real an imaginary parts,
\begin{equation}
  V_a=\mathcal{R}_a+i\, \mathcal{I}_a=\begin{pmatrix} x^I \\y_I\end{pmatrix}+i\begin{pmatrix} u^I \\v_I\end{pmatrix} \;,
\end{equation}
where 
\begin{equation}
\begin{pmatrix} u^I \\v_I\end{pmatrix} = \begin{pmatrix} \frac{\partial H}{\partial y_I}  \\ -
\frac{\partial H}{\partial x^I} 
\end{pmatrix} \;.
\label{eq:u-v-ref} 
\end{equation}
We rewrite \eqref{eq:u-v-ref} as
\begin{equation}
\mathcal{I}_a= \omega_{ab} \,  H_b \;,
\end{equation}
where $\omega_{ab}$ denotes the symplectic matrix
\begin{equation}
\omega=\begin{pmatrix}0&\mathbb{I}
\\-\mathbb{I} & 0   \end{pmatrix},
\label{omeg-symp}
\end{equation}
and $H_a$ denotes
\begin{equation}
H_a=\begin{pmatrix} \frac{\partial H}{\partial x^I}
\\ \frac{\partial H}{\partial y_I}   \end{pmatrix}.
\end{equation}
Using these, we obtain for \eqref{A-conn-def},
\begin{equation}
\mathscr{A}_{\mu} = \left(\omega {\cal R} \right)_a \, \partial_{\mu} {\cal R}_a + 
\left(\omega H \right)_a \, \partial_{\mu} H_a \;,
\end{equation}
and for the 
sigma-model Lagrangian \eqref{lag-sig-def}, 
\begin{equation}\label{sig}
 L_{\sigma} =2\partial^{\mu}\mathcal{R}_a\partial_{\mu}H_a-
\left[\left(\omega\mathcal{R}\right)_a\partial_{\mu}\mathcal{R}_a+
\left(\omega H\right)_a\partial_{\mu}H_a
\right]^2.
\end{equation}
Here $H_a$ depends on the real coordinates ${\cal R}$ and on the chiral background field $\Upsilon$
and $\bar \Upsilon$.  

Now we consider the case when the chiral background field $\Upsilon$ is taken to be constant.
Then we have the relation
\begin{eqnarray}
\partial_{\mu}H_a=H_{ab} \, \partial_{\mu}\mathcal{R}_b \;,
\label{HI-rel}
\end{eqnarray}
where $H_{ab}$ denotes the Hessian matrix
\begin{eqnarray}
H_{ab} = \begin{pmatrix}
\frac{\partial^2 H}{\partial x^I \partial x^J}  & \frac{\partial^2 H}{\partial y_I \partial x^J}  \\
\frac{\partial^2 H}{\partial x^I \partial y_J}  & \frac{\partial^2 H}{\partial y_I \partial y_J}
\end{pmatrix} \;.
\label{hessian}
\end{eqnarray}
Inserting \eqref{HI-rel} into \eqref{sig} yields
\begin{equation}
 L_{\sigma} =\Big[2H_{ab}-
\Big(\left(\omega\mathcal{R}\right)_a +
\left(\omega H\right)_c \, H_{ca} \Big)
\Big(\left(\omega\mathcal{R}\right)_b +
\left(\omega H\right)_d H_{db} \Big) \Big]
\partial^\mu \mathcal{R}_a \, \partial_\mu \mathcal{R}_b \;.
\label{sigma-real-hesse}
\end{equation}
In the rigid case, only the first term is present, which is encoded in the Hessian $H_{ab}$.
Thus, $H$ given in \eqref{eq:L-aux} serves as the Hesse potential for real special geometry
in the presence of a chiral background.  

Next, we compute the Hessian matrix $H_{ab}$.
Using \eqref{eq:u-v-ref} we express the Hessian as
\begin{eqnarray}
H_{ab} = \begin{pmatrix}
- \frac{\partial v_J}{\partial x^I }\big{|}_y  & - \frac{\partial v_J}{\partial y_I }\big{|}_x
 \\
\frac{\partial u^J}{\partial x^I }\big{|}_y  & \frac{\partial u^J}{\partial y_I }\big{|}_x
\end{pmatrix} \;,
\end{eqnarray}
and infer 
\begin{align}
\frac{\partial u^J}{\partial x^I }\big{|}_y =&\, - \frac{\partial v_I}{\partial y_J }\big{|}_x \;, \nonumber\\
\frac{\partial v_J}{\partial x^I }\big{|}_y = &\, \frac{\partial v_I}{\partial x^J }\big{|}_y \;, \nonumber\\
\frac{\partial u^J}{\partial y_I }\big{|}_x = &\, \frac{\partial u^I}{\partial y_J }\big{|}_x \;.
\end{align}
We proceed to compute these quantities.  To this end, we 
introduce the combinations
\begin{align}
R_{IJ} = &\, F_{IJ} + \bar{F}_{\bar I \bar J} \;, \nonumber\\
N_{IJ} = &\, - i \left(F_{IJ} -  \bar{F}_{\bar I \bar J} \right) \;,
\end{align}
as well as 
\begin{align}
[R_{\pm}]_{IJ} =&\,  R_{IJ} \pm 2 \, {\rm Re} \, F_{I \bar J} \;, \nonumber\\
[N_{\pm}]_{IJ} &=\, N_{IJ} \pm 2  \, {\rm Im} \, F_{I \bar J} \;.
\end{align}
Observe that
\begin{align}
[R_{\pm}]^T =&\,  R_{\mp} \;, \nonumber\\
[N_{\pm}]^T = &\, N_{\pm} \;.
\label{eq:symm-rel} 
\end{align}

We compute the Jacobian $J$ associated with the change of variables $(Y, \bar Y) \rightarrow (x, y)$,
\begin{align}
J = \frac{\partial(x, y)}{\partial(Y, \bar Y)} = 
\tfrac12 \left( \begin{array}{ccc}
\delta_J^I& \delta_J^I  \\
F_{IJ}+\bar F_{\bar I J} &  F_{I\bar J}+\bar F_{\bar I \bar J} \\
 \end{array} \right).
\end{align}
The inverse Jacobian is given by
\begin{equation}
J^{-1} = \frac{\partial(Y, \bar Y)}{\partial(x,y)} 
= \left(\begin{array}{ccc}
\delta^J{}_K+i\frac{\partial u^J}{\partial x^K} \big{|}_y & i\frac{\partial u^J}{\partial y_K}\big{|}_{x} \\
\delta^J{}_K-i\frac{\partial u^J}{\partial x^K}\big{|}_{y} & -i\frac{\partial u^J}{\partial y_K}\big{|}_{x}  \\
 \end{array} \right).
\end{equation}
Then, by imposing $J^{-1}J=\mathbb{I}$, we obtain the following relations, 
\begin{equation}
\frac{\partial u^I}{\partial x^J}\big{|}_{y}= [N_-^{-1} \, R_+]^I{}_{J}
\end{equation}
\begin{equation}
 \frac{\partial u^I}{\partial y_J}\big{|}_{x}=-2 \, [N_-^{-1}]^{IJ} \;.
\end{equation}
Using \eqref{eq:darboux}, 
we compute
\begin{align}
\frac{\partial v_I}{\partial x^J}\big{|}_u =&\, \tfrac12 \, [N_+]_{IJ} \;, \nonumber\\
\frac{\partial v_I}{\partial u^J}\big{|}_x =&\, \tfrac12 \, [R_-]_{IJ} \;, \nonumber\\
\end{align}
and obtain
\begin{align}
\frac{\partial v_I}{\partial x^J}\big{|}_y = &\, 
\frac{\partial v_I}{\partial x^J}\big{|}_u + 
\frac{\partial v_I}{\partial u^K}\big{|}_x \frac{\partial u^K}{\partial x^J}\big{|}_y 
\nonumber\\
=&\, \tfrac12 \, [N_+ + R_- \, N^{-1}_- \, R_+ ]_{IJ} \;.
\end{align}
Observe that this relation is symmetric by virtue of \eqref{eq:symm-rel}, as it should.

Using these relations, we obtain 
\begin{align}
H_{ab} = &\,
\begin{pmatrix}
- \tfrac12 \, [N_+ + R_- \, N^{-1}_- \, R_+ ]_{IJ}
& [N_-^{-1} \, R_+]^I{}_{J} \\
 [N_-^{-1} \, R_+]^J{}_{I}
 & -2 \, [N_-^{-1}]^{IJ}
\end{pmatrix} \nonumber\\
= &\, \begin{pmatrix}
- \tfrac12 \, [N_+ + R_- \, N^{-1}_- \, R_+ ]_{IJ}
& [N_-^{-1} \, R_+]^I{}_{J} \\
 [R_- \, N_-^{-1}]_I{}^{J}
 & -2 \, [N_-^{-1}]^{IJ}
\end{pmatrix}
\;.
\end{align}
This expresses the Hessian in terms of second derivatives of the function $F(Y, \bar Y, \Upsilon, \bar \Upsilon)$.
In the absence of a chiral background, this reduces to the expression found in \cite{Mohaupt:2011aa}.
Using \eqref{omeg-symp}
we obtain the relation
\begin{align}
H \, \omega \, H = \omega + \sigma \;\;\;,\;\;\;
\sigma = &\, \begin{pmatrix}
A  & B\\
C & D
\end{pmatrix} \;,
\label{om-si-om}
\end{align}
where
\begin{align}
A = &\,  - \tfrac12 \left(N_+ \, N_-^{-1} \, R_+ - R_- \, N_-^{-1} \, N_+ + R_-  \, N_-^{-1} \left(R_+ - R_- \right)
N_-^{-1} \, R_+ \right) = - A^T \;, \nonumber\\
B = &\, \left( N_+ - N_- \right) N_-^{-1} + R_- \, N_-^{-1} \left(R_+ - R_- \right) N_-^{-1} \;,
\nonumber\\
C = &\, - B^T \;, \nonumber\\
D = &\, - 2 \, N_-^{-1} \left(R_+ - R_- \right) N_-^{-1} = - D^T\;.
\label{eq:abcd}
\end{align}
This yields $\sigma = - \sigma^T$, as expected, since $[H \, \omega \, H ]^T = - H \, \omega \, H$.
Observe that 
\begin{equation}
\sigma = 0 \leftrightarrow F_{I \bar J } = 0 \;.
\end{equation}

The Hesse potential $H$ is homogeneous of degree two under real
rescalings of $\cal{R}$ and of $\Upsilon$, i.e.
$H(\lambda {\cal R}, \lambda^w \Upsilon, \lambda^w \bar \Upsilon) = \lambda^2 H({\cal R}, 
\Upsilon, \bar \Upsilon)$.  In the absence of a chiral background this yields the relation
$H_a = H_{ab} {\cal R}_b$ which, when combined with $H \, \omega \, H = \omega$ (which follows from 
\eqref{om-si-om}), results in $\left(\omega H\right)_c \, H_{ca} = \left( \omega {\cal R} \right)_a$.
This in turn leads to a simplification of \eqref{sigma-real-hesse}, namely to a 
doubling of the coefficient of the term $\left( \omega {\cal R} \right)_a$.

We proceed to show that the field strength \eqref{field-A} is also expressed in terms of the quantities in 
\eqref{eq:abcd}.
Using \eqref{eq:darboux} we get for the field strength 2-form, 
\begin{equation}
\mathscr{F}
= - \left(d {\bar Y}^{I} \wedge d F_I + d Y^I \wedge d {\bar F}_{\bar I} \right) 
= - 2 \left( dx^I \wedge dy_I + du^I \wedge dv_I \right) \;.
\end{equation}
We compute
\begin{eqnarray}
2 du^I \wedge d v_I &=& \left( \frac{\partial u^I}{\partial x^J} \Big{|}_{y} 
\frac{\partial v_I}{\partial x^K}  \Big{|}_{y} 
- (J \leftrightarrow K) \right) dx^J \wedge dx^K \nonumber\\
&& + 2  \left( \frac{\partial u^I}{\partial x^J} \Big{|}_{y} 
\frac{\partial v_I}{\partial y_K}  \Big{|}_{x} 
- 
\frac{\partial v_I}{\partial x^J} \Big{|}_{y} 
\frac{\partial u^I}{\partial y_K}  \Big{|}_{x} 
\right) dx^J \wedge dy_K \nonumber\\
&&
+ \left( \frac{\partial u^I}{\partial y_J} \Big{|}_{x} 
\frac{\partial v_I}{\partial y_K}  \Big{|}_{x} 
- (J \leftrightarrow K) \right) dy_J \wedge dy_K \;,
\end{eqnarray}
and obtain
\begin{eqnarray}
\mathscr{F} = - 2 \left( 2 \, \delta^K{}_J - C^K{}_J \right) \,  dx^J \wedge d y_K - A_{JK}  dx^J \wedge dx^K
- D^{JK} dy_J \wedge dy_K\;,
\label{eq:U1-F}
\end{eqnarray}
where the matrices $A, C$ and $D$ are given in \eqref{eq:abcd}. When $\sigma =0$ we get
$\mathscr{F} = - 4 \, dx^J \wedge d y_J$, as expected.

In \cite{Mohaupt:2011aa} a new formulation of the local c-map was given by making use of the real formulation
of special geometry in terms of the Hesse potential.  It would be interesting to extend their analysis to the
case with a chiral background.

\subsection*{Acknowledgements}

We thank B. de Wit and S. Mahapatra for discussions.
The work of G.L.C. was
partially supported by the Center for Mathematical Analysis, Geometry
and Dynamical Systems (IST/Portugal), as well as 
by Funda\c{c}\~{a}o para a Ci\^{e}ncia e a Tecnologia
(FCT/Portugal) through grant PTDC/MAT/119689/2010.  
The work of A.V.-O. was supported by Funda\c{c}\~{a}o para a Ci\^{e}ncia e a Tecnologia
(FCT/Portugal) through grant
SFRH/BD/64446/2009.

%%%%%%%%%%%%%%%%%%%%%%%%%%%%%%%%%%%%%%%%%%%%%%%%%%%%%%%
%%%%%%%%%%%%%%%%%%%%%%%%%%%%%%%%%%%%%%%%%%%%%%%

\begin{appendix}

\section{Derivatives \label{sec:derivatives}}

The first-order derivatives of $\Delta(z, \bar z, \Psi, \bar \Psi, R)$ and of 
$F(Y, \bar Y, \Upsilon, \bar \Upsilon)= \left( Y^0 \right)^2 \, {\cal F}(z, {\bar z}, \Psi, \bar \Psi, R) $ are given by
\begin{eqnarray}
\Delta_i &=& {\cal F}_i - \left(z^j - {\bar z}^{j} \right) {\cal F}_{ji}  - w \,\Psi {\cal F}_{\Psi i }
- R \,  {\cal F}_{R i } \;,
\nonumber\\
\Delta_{\bar \imath} &=& 2 {\cal F}_{\bar \imath} + {\cal F}_i - \left(z^j - {\bar z}^{j} \right) {\cal F}_{j\bar \imath}  - w \,\Psi {\cal F}_{\Psi \bar \imath }
- R \,  {\cal F}_{R \bar \imath } \;,\nonumber\\
\Delta_{\Psi} &=& (2 - w) \,{\cal F}_{\Psi} - \left(z^i - {\bar z}^{i} \right) {\cal F}_{i \Psi}
- w\, \Psi \, {\cal F}_{\Psi \Psi} -
R \,  {\cal F}_{R \Psi } \;,\nonumber\\
\Delta_{\bar \Psi} &=& 2 \,{\cal F}_{\bar \Psi} - \left(z^i - {\bar z}^{i} \right) {\cal F}_{i \bar \Psi}
- w\, \Psi \, {\cal F}_{\Psi \bar \Psi} -
R \,  {\cal F}_{R \bar \Psi } \;,\nonumber\\
\Delta_R &=& {\cal F}_{R} - \left(z^i - {\bar z}^{i} \right) {\cal F}_{i R}
- w \, \Psi \, {\cal F}_{\Psi R} -
R \,  {\cal F}_{R R } \;,
\label{der1-delta}
\end{eqnarray}
and by 
\begin{eqnarray}
F_0&=&Y^0(2\mathcal{F}-z^i\mathcal{F}_i - w   \Psi \mathcal{F}_ {\Psi} - R \mathcal{F}_{R}) \;,
\nonumber\\
F_{\bar 0}&=& Y^0 \left({\cal F}_R - {\bar R} \,
{\bar z}^i \, \mathcal{F}_{\bar \imath} - w \, {\bar R} \, {\bar \Psi} \,  \mathcal{F}_{\bar \Psi} \right) \;,
\nonumber\\
F_i&=&Y^0\mathcal{F}_i  \;,\nonumber\\
F_{\bar \imath}&=&Y^0  {\bar R} \, \mathcal{F}_{\bar \imath}  \;,\nonumber\\
F_{\Upsilon}&=&\left(Y^0\right)^{2-w}\mathcal{F}_\Psi \;,
\nonumber\\
F_{\bar \Upsilon}&=&{\bar R}^2\left({\bar Y}^0\right)^{2-w} {\cal F}_{\bar \Psi} \;, 
\label{der1-F}
\end{eqnarray}
respectively.
The second-order derivatives of $F(Y, \bar Y, \Upsilon, \bar \Upsilon)$ read,
\begin{eqnarray}
F_{00}&=&2\mathcal F-2z^i\mathcal F_ i+z^iz^j\mathcal F_{ij}
+ w (w-3) \,
\Psi\mathcal F_{\Psi}+ 2 w \, z^i\Psi\mathcal F_{i\Psi}+ w^2 \, \Psi^2 \mathcal F_{\Psi\Psi}
\nonumber\\
&&-2R\mathcal F_R+2z^iR\mathcal F_{iR}+2w R \Psi \mathcal F_{R\Psi}+R^2\mathcal F_{RR}
\nonumber\\
F_{0\bar0}&=& \Delta_{\Psi} - {\bar z}^i \, {\cal F}_{i \Psi} 
\nonumber\\
&&
+ \bar R \left( z^j\bar z^{i}\mathcal F_{\bar \imath j}-2\bar z^{i}\mathcal{F}_{\bar \imath}
-2 w\bar \Psi\mathcal{F}_{\overline{\Psi}} + w R \bar \Psi  {\cal F}_{R \bar \Psi } +  {\bar z}^i R
{\cal F}_{\bar \imath R} \right. \nonumber\\
&& \left. \qquad +w z^i  \overline{\Psi}\mathcal{F}_{i \overline{\Psi}} +
w {\bar z}^i 
\Psi\mathcal{F}_{\bar \imath \Psi}+w^2 \, \overline{\Psi}\Psi
\mathcal{F}_{\overline{\Psi}\Psi}\right) \;,
\nonumber\\
F_{0i}&=&\Delta_i - {\bar z}^j \, {\cal F}_{ji}
\nonumber\\
F_{0\bar \imath}&=&\bar R\left(\Delta_{\bar \imath} - {\cal F}_i - {\bar z}^j \, {\cal F}_{j \bar \imath}
\right)
\nonumber\\
F_{i \bar 0} &=& {\cal F}_{R i} - {\bar R} {\bar z}^j {\cal F}_{\bar \jmath i } - w {\bar R} {\bar \Psi} 
{\cal F}_{\bar \Psi i} \;, \nonumber\\
F_{0 \Upsilon} &=& \left(Y^0\right)^{1-w} \left(\Delta_{\Psi} - {\bar z}^i  {\cal F}_{i \Psi} \right)
\;,  \nonumber\\
F_{0 \bar \Upsilon} &=& {\bar R} \left(\bar Y^0\right)^{1-w} \left(\Delta_{\bar \Psi} - {\bar z}^i  {\cal F}_{i \bar \Psi} \right)
\;,  \nonumber\\
F_{ij} & = & {\cal F}_{ij} \;, \nonumber\\
F_{i\bar \jmath}&=& \bar R\mathcal{F}_{i \bar \jmath} \nonumber\\
F_{i \Upsilon} &=& \left(Y^0\right)^{1-w} \, {\cal F}_{i \Psi} \;, \nonumber\\
F_{i \bar \Upsilon} &=& {\bar R} \, \left( \bar{Y}^0\right)^{1-w} \, {\cal F}_{i \bar \Psi} \;.
\label{der2-F}
\end{eqnarray}
In \eqref{der1-F} and \eqref{der2-F}, 
the derivatives on the left hand side are with respect to the fields $(Y^I, \bar{Y}^{\bar I}, \Upsilon, 
\bar \Upsilon)$, while the derivatives on the right hand side refer to the projective coordinates 
\eqref{project-coord} and their complex conjugates.

\end{appendix}

%\bibliographystyle{JHEP}
%\bibliography{references}

\providecommand{\href}[2]{#2}\begingroup\raggedright\endgroup

\end{document}